\documentclass[onecolumn,12pt,draftcls]{IEEEtran}
\IEEEoverridecommandlockouts

\usepackage{geometry}
\def\BibTeX{{\rm B\kern-.05em{\sc i\kern-.025em b}\kern-.08em T\kern-.1667em\lower.7ex\hbox{E}\kern-.125emX}}
\usepackage{amsmath}
\usepackage{subfigure}
\usepackage{amsfonts}
\usepackage{bbding}
\usepackage{amssymb}
\usepackage{array}
\usepackage{multirow}

\usepackage{graphicx}
\usepackage{algorithmic}
\usepackage[compress]{cite}
\makeatletter
\renewcommand{\citepunct}{,\penalty\@m\hskip.13emplus.1emminus.1em}
\renewcommand{\citedash}{\hbox{--}\penalty\@m}
\makeatother
\usepackage{color}
\allowdisplaybreaks

\usepackage{amsthm}
\usepackage{stfloats}

\newtheorem{condition}{Condition}

\newtheorem{rem}{Remark}

\newtheorem{prop}{Proposition}

\newcommand{\blue}[1]{{\textcolor[rgb]{0,0,1}{#1}}}

\begin{document}
\title{Deep Learning for Radio Resource Allocation with Diverse Quality-of-Service Requirements in 5G}

\author{
	\IEEEauthorblockN{Rui Dong,~\IEEEmembership{Student Member,~IEEE,} Changyang She,~\IEEEmembership{Member,~IEEE,}
	Wibowo~Hardjawana,~\IEEEmembership{Member,~IEEE,}		
	Yonghui~Li,~\IEEEmembership{Fellow,~IEEE,}			
	and~Branka~Vucetic,~\IEEEmembership{Fellow,~IEEE}}
	\thanks{This paper has been presented in part at the IEEE Global Communications Conference 2019 \cite{Changyang2019GC}.}
	\thanks{The authors are with the School of Electrical and Information Engineering, University of Sydney, Sydney, NSW 2006, Australia (email: \{rui.dong, changyang.she, wibowo.hardjawana, yonghui.li, branka.vucetic\}@sydney.edu.au).}
%	\thanks{Manuscript received January 16, 2019; revised May 07, 2019 and July 01, 2019; accepted July 02, 2019. The associate editor coordinating the review of this paper and approving it for publication was Prof. C. Zhong. \textit{(Corresponding author: Changyang She.)}}
%	\thanks{The work of W. Hardjawana was supported by Australian Research Council Discovery Early Career Research Award DE150101704. The work of Y. Li was supported by Australian Research Council under Grants DP150104019 and DP190101988. The work of B. Vucetic was supported by the Australian Research Council Laureate Fellowship under Grant FL160100032. }
}

\maketitle

\begin{abstract}
To accommodate diverse Quality-of-Service (QoS) requirements in 5th generation cellular networks, base stations need real-time optimization of radio resources in time-varying network conditions. This brings high computing overheads and long processing delays. In this work, we develop a deep learning framework to approximate the optimal resource allocation policy that minimizes the total power consumption of a base station by optimizing bandwidth and transmit power allocation. We find that a fully-connected neural network (NN) cannot fully guarantee the QoS requirements due to the approximation errors and quantization errors of the numbers of subcarriers. To tackle this problem, we propose a cascaded structure of NNs, where the first NN approximates the optimal bandwidth allocation, and the second NN outputs the transmit power required to satisfy the QoS requirement with given bandwidth allocation. Considering that the distribution of wireless channels and the types of services in the wireless networks are non-stationary, we apply deep transfer learning to update NNs in non-stationary wireless networks. Simulation results validate that the cascaded NNs outperform the fully connected NN in terms of QoS guarantee. In addition, deep transfer learning can reduce the number of training samples required to train the NNs remarkably.
\end{abstract}


%\IEEEpeerreviewmaketitle

\section{Introduction}
\subsection{Background}
The 5th Generation (5G) cellular networks are expected to support various emerging applications with diverse Quality-of-Service (QoS) requirements, such as enhanced mobile broadband services, massive machine-type communications, and Ultra-Reliable and Low-Latency Communications (URLLC) \cite{3GPP2017Scenarios}. To guarantee the QoS requirements of different types of services, existing optimization algorithms for radio resource allocation are designed to maximize spectrum efficiency or energy efficiency by optimizing scarce radio resources, such as time-frequency resource blocks and transmit power, subject to QoS constraints \cite{3GPP2017EE,zhao2016cluster,buzzi2016survey,6G2019EE,xu2016energy,sun2019optimizing,amjad2019effective}.

There are two major challenges for implementing existing optimization algorithms in practical 5G networks. First, QoS constraints of some services, such as delay-sensitive and URLLC services, may not have closed-form expressions. To execute an optimization algorithm, the system needs to evaluate the QoS achieved by a certain policy via extensive simulations or experiments, and thus suffers from long processing delay \cite{amjad2019effective,hu2018optimal}. Second, even if the closed-form expressions of QoS constraints can be obtained in some scenarios, the optimization problems are non-convex in general \cite{hu2018optimal,ye2019power,sun2019optimizing}. The system also needs to update resource allocation by solving non-convex problems to accommodate the time-varying channel and traffic conditions, leading to very high computing overhead. Even for some convex optimization problems that can be solved by well-developed methods, like the interior-point method, the computing complexity is still too high {to be implemented in real time} \cite{boyd}.

%The improvement of energy efficiency is a crucial goal for supporting different kinds of services in wireless networks \cite{6G2019EE}.
%Resource management is widely adopted to achieve the goal of energy-saving \cite{Buzzi2016EE,She2017URLLCpower}. Optimization methods are usually used to obtain global optimal solutions for resource allocation subject to the diverse QoS requirements of different services.
%However, optimizing resource management problems over hybrid 5G services are general mixed-integer non-linear programming (MINLP) \cite{hu2018optimal,ye2019power}, it is very challenging to obtain global optimal solutions with low complexity. In addition to it, due to the stringent delay requirements, such as delay-sensitive and URLLC services, the implementation period of optimization methods is another issue for solving optimization problems in real-time. A low-complexity algorithm subject to QoS can be a potential approach to solve a problem in real-time  when the problem is of small-scale \red{(add ref our globlecom paper)}.

Deep learning is a promising approach to find the optimal resource allocation in real time \cite{Sun2018dnnInterference,zappone2018online,zappone2018modelaided,liang2018dnnPowerContr,lei2019learning}. The basic idea is to use an artificial Neural Network (NN) to approximate the optimal resource allocation policy that maps the system states to the optimal resource allocation. The system first trains the NN off-line with a large number of labeled samples. After the training phase, the optimal resource allocation can be obtained from the output of the NN for any given input. According to the Universal Approximation Theory, if the optimal policy is a deterministic and continuous function, then the approximation {errors approach to zero as the number of neurons goes to infinite} \cite{HORNIK1989UniversalApprox}.

It is worth noting that the application of deep learning in wireless networks is not straightforward.
For some discrete optimization variables, such as the number of subcarriers, antennas and the user association decisions, the approximation of the NN can be inaccurate due to the quantization of these discrete variables. As a result, the solution obtained from the NN cannot fully guarantee the QoS requirements of different types of services. In addition, deep learning requires a large number of labeled training samples. To obtain labeled training samples, we should first design an optimization algorithm to solve the formulated optimization problem. Even if a large number of labeled training samples are obtained with the optimization algorithm, the pre-trained NN is not accurate when the wireless network is non-stationary. For example, the distribution of wireless channels and the types of services in the network may vary. These non-stationary parameters that are not included in the input of the NN are referred to as hidden variables \cite{riley2019three}. During the training phase, we assume that the hidden variables are fixed. However, in practical systems, these hidden variables drift over time. As discussed in \cite{riley2019three}, the dynamic hidden variables can be pernicious in deep learning.

\subsection{Related Works}
Improving resource utilization efficiency for different kinds of services has been extensively studied in the literature. For delay-tolerant services, the QoS requirement is formulated as an average data rate requirement in Orthogonal Frequency Division Multiple Access (OFDMA) systems \cite{Zhikun2013Energy}, where the subcarrier and transmit power allocation and antenna configuration were optimized. To guarantee the queueing delay bound and the delay bound violation probability of real-time services, effective capacity was adopted in \cite{EEECLY,yu2017statistical} to optimize bandwidth allocation and power control schemes. In URLLC, to reduce transmission delay, the blocklength of channel codes is short, and the fundamental relation between decoding error probability and blocklength was derived in \cite{Yury2014Quasi}. This relation was used to optimize resource allocation for short packet transmissions in URLLC \cite{xu2016energy,zhu2019energy,sun2019optimizing}. For most of these problems, the QoS constraints do not have closed-form expressions and the optimization algorithms cannot be executed in a real-time manner.

Approximating optimal resource allocation policies with NNs has been investigated in \cite{Sun2018dnnInterference,lee2018PowerParallel,zappone2018online}. The authors of \cite{Sun2018dnnInterference} proved that an iterative algorithm for power control in wireless networks can be accurately approximated by a Fully-connected Neural Network (FNN). In \cite{lee2018PowerParallel} and \cite{lei2019learning}, convolutional neural networks were used to approximate the power control policy and the content delivery policy, respectively. To improve energy efficiency, \cite{zappone2018online} proposed an online deep learning approach to approximate the energy-efficient power control scheme obtained from the monotonic fractional programming framework in \cite{zappone2017globally}. When the optimal optimization algorithm is not available, unsupervised deep learning {was applied in \cite{Eisen2018Constraint,lee2019deep}, where} the parameters of a NN are trained to satisfy the Karush-Kuhn-Tucker (KKT) conditions of the optimization problem. However, for problems with integer variables that are not defined over a compact set, the KKT conditions do not exist.

Considering that wireless networks are highly dynamic, NNs trained offline {cannot} achieve good performance in non-stationary networks. To handle this issue, deep transfer learning was used in some existing works. For example, when data arrival processes \cite{Chen2019ResMagm}, traffic patterns \cite{Zhang2019TrafficPred}, or the size of the network \cite{Yao2019TF,CHATURVEDI2015IntroTF} change, deep transfer learning can be used to fine-tune the pre-trained NNs. %Nevertheless, how to guarantee diverse QoS requirements in beyond 5G systems deserves further study.

\subsection{Our Contributions}
Motivated by the above issues, we will answer the following questions in this paper: 1) How to design an optimization algorithm that can find the optimal resource allocation subject to diverse QoS requirements?
2) How to improve the approximation accuracy of the NN when there are quantization errors of discrete optimization variables? 3) How to adapt the pre-trained NN according to non-stationary wireless networks? To illustrate our approach, we consider an example problem that minimizes the total power consumption. The method can be easily extended to other kinds of problems, such as maximizing spectrum efficiency. Our main contributions are summarized as below:
\begin{itemize}
\item We establish a deep learning framework that can obtain a near-optimal energy-efficient bandwidth and transmit power allocation scheme in 5G New Radio (NR) systems, where the QoS requirements of delay-tolerant, delay-sensitive, and URLLC services are satisfied. The optimization problem is a Mixed Integer Non-Linear Programming (MINLP) since the number of subcarriers allocated to each user is an integer and the transmit power is a continuous variable.
\item To obtain training samples, we develop an optimization algorithm to solve the MINLP, and analyze the convergence conditions, in which the algorithm converges to the global optimal solution of the MINLP. In addition, we prove that the conditions hold for delay-tolerant and delay-sensitive services. For URLLC, our analysis shows that the conditions hold in an asymptotic scenario, where the number of antennas is sufficiently large. Our numerical results validate that the conditions also hold in non-asymptotic scenarios.
\item We observe that the output of an FNN cannot guarantee the QoS requirement of different types of services. To address this issue, we develop a cascaded structure of NNs. The first NN obtains bandwidth allocation for multiple users. Given bandwidth allocation, the transmit power that is required to satisfy the QoS requirement of each user is obtained from the second NN.
\item We adopt deep transfer learning to fine-tune pre-trained NNs in non-stationary wireless networks. The basic idea is to reuse the first several layers of the pre-trained NNs and train the last a few layers with a small number of new training samples. Numerical and simulation results show that the cascaded NNs can converge quickly in non-station wireless networks.
%
%	We show how to collect training samples to train DNN for an MINLP problem of minimizing the total power consumption subject to QoS requirements. We propose a low-complexity algorithm to optimize resources of the transmit power and subcarriers. The optimality conditions of the algorithm are provided. Theoretical analysis shows that the conditions hold for both delay-tolerant and delay-sensitive services. For URLLC, the conditions hold when the number of antennas at the base station (BS) is large. Numerical results validate that the conditions hold for URLLC even when the number of antennas is small.
	
	%single DNN, to relieve the cost of training samples collection,  The proposed framework also extends to multiple services, where the resource allocation policy can learn from multiple DNNs to satisfy the increasing number of services. This framework achieves near-optimal performance with the requirement of only a small amount of training samples.
%	
%	\item
%	Numerical results validate the superiority of the proposed framework in terms of the number of training samples and performance accuracy. The scalability and robustness are also evaluated and demonstrated for our deep transfer learning framework.
\end{itemize}

The rest of the paper is organized as follows. In Section II, we formulate the system models. The cascaded NNs for ensuring the QoS requirement are presented in Section III. In Section IV, we apply deep transfer learning in non-stationary wireless networks. We provide simulation results in Section V and conclude the work in Section VI. All the notations used in this chapter are listed in Table \ref{ch5:notations}.

\begin{table*}[htb]
	\centering
	\caption{Notations}
	\label{ch5:notations}
\vspace{-0.2cm}	
	\scalebox{1.0}{
		\begin{tabular}{|p{1.8cm}| p{5.5cm} ||p{1.0cm}| p{7.5cm}|}%{|l|l||l|l|}
			\hline
			Notation & Definition  & Notation & Definition \\ \hline
		    $(\cdot)^{\rm T}$ & transpose operator & $K$ & total number of users \\ \hline
			$\xi \in \{\rm t, s, u\}$ & superscript representing delay-tolerant, delay-sensitive and URLLC services & $\mathcal{K}^{\xi}$& set of users \\ \hline
			$W$ & bandwidth of each subcarrier & $T_s$  &  duration of each slot  \\ \hline
			$T_c$ & channel coherence time & $D_k^{\rm q, s}$ & delay bound of the $k$-th delay-sensitive service\\ \hline
			$N_{\rm T}$ & number of antennas at the BS & $\epsilon_k^{\rm q,s}$ & maximal tolerable delay bound violation probability \\ \hline
			$\epsilon^{\rm max, u}$ & threshold of decoding error probability & $\alpha_k^{\xi}$ & large-scale channel gain of the $k$-th user\\ \hline
			$g_{k,n}^\xi$ & small-scale channel gain on the $n$-th subchannel of the $k$-th user& $P_k^\xi$ & transmit power allocated to the $k$-th user \\ \hline
			$N_0$ & single-side noise spectral density & $\bar{a}_k$ & average data arrival rate of the $k$-th delay-tolerant user\\ \hline
			$N_{k}^\xi$ & number of allocated subcarriers & $\theta_k^{\rm s}$ & QoS exponent of the $k$-th delay-sensitive service  \\ \hline
			$\nu^{\rm s}$ & inverse of average packet size of delay-sensitive services & $\nu^{\rm a}$ & average inter-arrival time between packets of delay-sensitive services \\ \hline
			$\epsilon_k^{\rm d, u}$ & decoding error probability of the $k$-th URLLC user & $B_k^{\rm u}$ & number of bits in each packet of the $k$-th URLLC user  \\ \hline
			$V_k^{\rm u}$ & channel dispersion of the $k$-th URLLC user& $\bar{\epsilon}_k^{\rm d, u}$ & average decoding error probability of the $k$-th URLLC user\\ \hline
			$\rho$ & power amplifier efficiency & $P^{\rm ca}$ & power consumption by each antennas \\ \hline
			$P_0^{\rm c}$ & fixed circuit power consumption & $c_k^{\xi}$ & feature of the packet arrival process of the $k$-th user \\ \hline
			
		\end{tabular}
	}
\end{table*}

%Deep transfer learning has emerged as a desirable technique to deal with the mismatch problem \cite{Pan2010TFsurvey}, which attempts to transfer the related knowledge from the source task to the target task to achieve the desired performance, at the meantime, significantly reducing the training time and the number of training samples.
%There are several deep transfer learning works to deal with the dynamics in wireless communications. In \cite{Chen2019ResMagm}, the dynamics of wireless communication was naturally defined by the transmission parameters across time. It was a straightforward concept to transfer from the knowledge of past successfully transmission experience to the future transmission decision.
%In \cite{Zhang2019TrafficPred}, the dynamics were defined as cellular traffic patterns. By clustering the geographical zones based on the different known traffic patterns, DNN was used to predict the future traffic pattern. Moreover, \cite{Yao2019TF,CHATURVEDI2015IntroTF} defined the dynamics by the size of the network, where a DNN learned high dimensional network from low dimensional forms via deep transfer learning. Nevertheless, how to apply deep transfer learning for allocating resources among different services with diverse QoS requirements remains unclear.

\section{System Model and Problem Formulation}
\label{sysmodel}

%{\color{red} \subsection{General Optimization Problem}
%In wireless networks, the optimization problem on resource management can be generalized as mixed integer non-linear programming problems, which consist of continuous optimization variables $\boldsymbol{p}$, such as transmit power, subject to some constraints, and discrete optimization variables $\boldsymbol{\beta}$, such as user association and subcarrier allocation. Such an optimization problem can be written as
%
%\begin{align}
%\label{general}
%&\min_{\boldsymbol{p},\boldsymbol{\beta}} f(\boldsymbol{p},\boldsymbol{\beta}) ,\\
%\text{s.t.} \;
%& g(\boldsymbol{p},\boldsymbol{\beta}) \leq  \boldsymbol{c}, \nonumber \\
%& p_k \in \mathbb{C}, \beta_k \in \mathbb{N} ,\nonumber
%\end{align}
%where $f(\cdot, \cdot)$ is the objective function, such as spectrum efficiency and energy efficiency, $g(\cdot, \cdot)$ is the constraint function subject to a constant variable $\boldsymbol{c}$, e.g., QoS requirement, subcarrier and power constraints, and $p_k$ and $\beta_k$ are the $k$-th element of $\boldsymbol{p}$ and $\boldsymbol{\beta}$.
%
%Next, we will provide a typical example of such a mixed integer non-linear programming problem, including resource allocation and power control.}

\subsection{System Model}
We consider a downlink OFDMA system, where one multi-antenna BS serves $K$ single-antenna users that request different kinds of services, including delay-tolerant, delay-sensitive and URLLC services. The corresponding sets of users are denoted by $\mathcal{K}^{\rm t}$, $\mathcal{K}^{\rm s}$, and $\mathcal{K}^{\rm u}$, respectively. For notational simplicity, we use a superscript $\xi \in \{\rm t, s, u\}$ to represent delay-tolerant, delay-sensitive and URLLC services. The bandwidth of each subcarrier and the duration of one Transmission Time Interval (TTI) in the OFDMA system are denoted by $W$ and $T_{\rm s}$, respectively.

\subsubsection{Channel Model} We assume that channels are block fading in both time and frequency domains, and the channel gains on different subcarriers allocated to one user are independent and identical distributed (i.i.d). Channel coherence time is denoted by $T_{\rm c}$, which is much longer than the duration of a TTI, $T_{\rm s}$. We consider downlink (DL) transmissions and assume that channel state information (CSI) is only available at users to avoid the overhead for channel estimations at the BS.

\subsubsection{Queueing Model} For all kinds of services, packets in the buffer of the BS are served according to the first-come-first-serve order. For delay-tolerant services, we only need to ensure the stability of the queueing system. For delay-sensitive services, a delay bound, $D_k^{\rm q, s}$, and a maximal tolerable delay bound violation probability, $\epsilon_k^{\rm q, s}$, should be satisfied. To avoid queueing delay for URLLC, packets should be served immediately after arriving at the BS. The decoding error probability of packets should not exceed a required threshold, $\epsilon^{\max,{\rm u}}$.

\subsection{Delay-Tolerant Services}
For delay-tolerant services, the blocklength of channel code can be sufficiently long, and the average data rate of each user approaches Shannon's capacity, i.e.,
\begin{equation}
\bar{R}_k^{\rm t}
= N_{k}^{\rm t} \mathbb{E}_{g_{k,n}^{\rm t}} \left[
 W \ln\left( 1+\frac{ \alpha_{k}^{\rm t} g_{k,n}^{\rm t} P_k^{{\rm t}}}{ N_0 {N_{\rm T}} N_{k}^{\rm t} W}\right)   \right] \;\text{(bits/s)},\label{rb}
\end{equation}
where $\alpha_{k}^{\rm t}$ is the large-scale channel gain, $g_{k,n}^{\rm t}$ is the small-scale channel gain on the $n$-th subchannel, $P_k^{ {\rm t}}$ is the transmit power, $N_0$ is the single-side noise spectral density, ${N_{\rm T}}$ is the number of antennas at the BS, and $N_{k}^{\rm t}$ is the number of subcarriers allocated to the $k$-th delay-tolerant user. Since CSI is not available at the BS, the transmit power is equally allocated on different antennas and subcarriers.

%\blue{Not here:} Through the paper, we consider the worst case of the propagation environment, i.e., there is no strong line-of-sight path between the BS and each of the users. In this case, the channels can be modelled as Rayleigh fading and the probability density function of small-scale channel gains follows

To ensure the stability of the queueing system, the average service rate should be equal to or higher than the average data arrival rate of the user, i.e.,
\begin{align}
\label{dtCon}
\bar{R}_k^{\rm t} \geq \bar{a}_k,
\end{align}
where $\bar{a}_k$ is the average data arrival rate of the $k$-th delay-tolerant user.
%This constraint can be written as in a form of $g(\boldsymbol{p},\boldsymbol{\beta}) \leq  \boldsymbol{c}$ in \eqref{general}.

\subsection{Delay-Sensitive Services}
%\subsubsection{Achievable Rate}
%For delay-sensitive services, the achievable rate can also be approximated by Shannon's capacity. The ergodic capacity  (bits/s) of the $k$th user, $\forall k \in \mathcal{K}^{\rm s}$, is similar to \eqref{rb}.
For delay-sensitive services, the blocklength of channel codes is finite. We denote $\Phi$ as the SNR gap between the channel capacity and a practical modulation and coding scheme as in \cite{Lingjia2012ICC,WirelessCom}. The value of $\Phi$ decreases with the blocklength of channel codes. For delay-tolerant services, the blocklength can be long enough such that $\Phi \to 1$. Thus, the achievable rate in \eqref{rb} is the channel capacity. For delay-sensitive services, due to the constraint on the transmission delay, the coding blocklength is finite, and thus $\Phi > 1$. The achievable rate of the $k$-th delay-sensitive user can be expressed as
\begin{align}
R_k^{{\rm{s}}} = \sum\limits_{n = 1}^{N_k^{\rm{s}}} {W{{\ln }}\left( {1 + \frac{{\alpha _k^{\rm{s}}g_{k,n}^{\rm{s}}P_k^{\rm{s}}}}{{\Phi {N_0}{N_{\rm{T}}}N_k^{\rm{s}}W}}} \right)},\;\text{(bits/s)}, \label{Rs}
\end{align}
where $\alpha_{k}^{\rm s}$ is the large-scale channel gain, $g_{k,n}^{\rm s}$ is the small-scale channel gain on the $n$-th subchannel, $P_k^{ {\rm s}}$ and $N_{k}^{\rm s}$ are the transmit power and the number of subcarriers allocated to the $k$-th delay-sensitive user, respectively.

To guarantee $D_k^{\rm q, s}$ and $\epsilon_k^{\rm q, s}$ for delay-sensitive services, effective bandwidth and effective capacity are widely used \cite{EB,EC}. We assume that the packet arrival process of each delay-sensitive user is a compound Poisson process\footnote{For some other kinds of packet arrival processes, the method to compute effective bandwidth can be found in \cite{Kelly1996EffBW,ozmen2015wireless}.}. The inter-arrival time between packets and the size of each packet follow exponential distributions with parameters $\nu^{\rm a}$ and $\nu^{\rm s}$, respectively. Then, the effective bandwidth of the $k$-th delay-sensitive user can be expressed as follows \cite{Kelly1996EffBW},
\begin{align}
\label{effB}
E_k^{\rm B, s} =
\frac{\nu^{\rm a}}{\nu^{\rm s} - \theta_k^{\rm s}},\; \text{(bits/s)},
\end{align}
where $\theta_k^{\rm s}$ is the QoS exponent, which can be obtained from
\begin{align}
\label{theta}
\exp\left[-\theta_k^{\rm s}{E_k^{\rm B, s}(\theta_k^{\rm s}) D_k^{\rm q, s}}\right] \approx \epsilon_k^{\rm q, s}.
\end{align}
Substituting \eqref{effB} into \eqref{theta}, we can derive that
\begin{align}
\label{thet}
\theta_k^{\rm s} =
\frac{\nu^{\rm s} \ln (\epsilon_k^{\rm q, s}) }
{\ln \epsilon_k^{\rm q, s} -\nu^{\rm a} D_k^{\rm q, s}}.
\end{align}
With the block fading channel model, $g_{k,n}^{\rm s}$ is constant within each block and is i.i.d. among different blocks. The duration of each block equals to the channel coherence time, $T_{\rm c}$. Thus, the effective capacity can be simplified as follows \cite{Wu2003EffCap,Tang2007QoS},
\begin{align}
&E_k^{\rm C, s} = - \frac{1}{\theta_k^{\rm s}{T_{\rm c}}} \ln \mathbb{E}_{g_{k,n}^{\rm s}} \left[ \exp \left(
- \theta_k^{\rm s}{T_{\rm c}} R_k^{\rm s}
\right)\right]\; \label{EC1}\\
& =  -\frac{N_k^{\rm s}}{\theta_k^{\rm s}{T_{\rm c}}}\ln\left[\mathbb{E}_{g_{k,n}^{\rm s}}\left(1+\frac{{\alpha _k^{\rm{s}}g_{k,n}^{\rm{s}}P_k^{\rm{s}}}}{{\Phi {N_0}{N_{\rm{T}}}N_k^{\rm{s}}W}}\right)^{-\varpi_k}\right] ,\; \text{(bits/s)}, \label{EC2}
\end{align}
where $\varpi_k = \frac{\theta^{\rm s}_kT_{\rm c}W}{\ln2}$, and \eqref{EC2} is obtained by substituting $R_k^{{\rm{s}}}$ in \eqref{Rs} into \eqref{EC1}. To guarantee $D_k^{\rm q, s}$ and $\epsilon_k^{\rm q, s}$, the following constraint should be satisfied \cite{liu2007resource},
\begin{align}
\label{dsCon}
E_k^{\rm C, s} \geq E_k^{\rm B, s}.
\end{align}
\begin{rem}
\emph{It is worth noting that the approximation in \eqref{theta} is accurate in the large delay regime. Since the delay requirement of delay-sensitive services is much longer than the channel coherence time, the queueing delay requirement can be satisfied with constraint in \eqref{dsCon} \cite{liu2007resource,Tang2007QoS}.}
\end{rem}

%This constraint can be written in a form of $g(\boldsymbol{p},\boldsymbol{\beta}) \leq  \boldsymbol{c}$ as in \eqref{general}, where $\theta_k^{\rm s}$ and $E_k^{\rm B, s}(\theta_k^{\rm s})$ can be directly calculated by \eqref{thet} and \eqref{effB}, respectively, thus, $E_k^{\rm B, s}(\theta_k^{\rm s})$ becomes a constant constraint, and $E_k^{\rm C, s}(\theta_k^{\rm s})$ is a function of $P_k^{\rm s}$ and $N_k^{\rm s}$.

\subsection{URLLC Services}
When transmitting short packets of URLLC, the blocklength of channel codes is much shorter than the previous services. The decoding errors in the short blocklength regime have a significant impact on the reliability of URLLC, and hence the decoding error probability should be considered in URLLC. Since \eqref{Rs} does not characterize the relationship between the decoding error probability and the achievable rate, it is not applicable for URLLC. According to the Normal Approximation of the achievable rate in the short blocklength regime in \cite{Yury2014Quasi} and the analysis in Appendix E of \cite{Changyang2018TWC}, the achievable rate over the frequency-selective channel can be approximated by
\begin{align}
\label{ru}
R_k^{\rm{u}} \approx \frac{W}{{\ln 2}}{\Bigg{\{}} \left[ {\sum\limits_{n = 1}^{N_k^{\rm{u}}} {\ln \left( {1 + \frac{{\alpha _k^{\rm{u}}g_{k,n}^{\rm{u}}P_k^{\rm{u}}}}{{{N_0}{N_{\rm{T}}}N_k^{\rm{u}}W}}} \right)} } \right]
- \sqrt {\frac{{{V^{\rm u}_k}}}{{{T_{\rm{s}}}W}}} f_Q^{ - 1}\left( {\epsilon _k^{{\rm{d,u}}}} \right) {\Bigg{\}}}, \; \text{(bits/s)},
\end{align}
where $\alpha_{k}^{\rm u}$ is the large-scale channel gain, $g_{k,n}^{\rm u}$ is the small-scale channel gain on the $n$-th subchannel, $P_k^{ {\rm u}}$ and $N_{k}^{\rm u}$ are the transmit power and the number of subcarriers allocated to the $k$-th URLLC user, respectively, ${\epsilon _k^{{\rm{d,u}}}}$ is the decoding error probability, $f_Q^{-1}$ is the inverse of Q-function, and $V_k^{\rm u}$ is the channel dispersion, which is given by $V_k^{\rm{u}} = N_k^{\rm{u}} - \sum\limits_{n = 1}^{N_k^{\rm{u}}} {\frac{1}{{{{\left( {1 + \frac{{\alpha _k^{\rm{u}}g_{k,n}^{\rm{u}}P_k^{\rm{u}}}}{{{N_0}{N_{\rm{T}}}N_k^{\rm{u}}W}}} \right)}^2}}}} $ \cite{Yury2014Quasi,Changyang2018TWC}.
According to the definition in \cite{Yury2010Channel}, the channel dispersion measures the stochastic variability of the channel relative to a deterministic channel with the same capacity.

The packet arrival process of a URLLC user can be modeled as a Bernoulli process, such as mission-critical IoT applications and vehicle networks \cite{3GPP2012MTC,Hassan2013A}. In other words, the number of packets arriving at the buffer of the BS in each TTI is either zero or one. To avoid queueing delay in the buffer of BS, the downlink transmission duration of a packet should be one TTI. Denote the number of bits in one packet as $B_k^{\rm u}$. From $T_{\rm s}R_k^{\rm{u}} = B_k^{\rm u}$, we can derive the average decoding error probability, i.e.,
\begin{align}
\bar{\epsilon}_k^{\rm d,u}
\approx \mathbb{E}_{g_{k,n}^{\rm u}}\Bigg\{ {f_Q}\Bigg( {\sqrt \frac{{{T_{\rm{s}}}W}}{{N_k^{\rm{u}}}}} \Bigg\{ \left[ {\sum\limits_{n = 1}^{N_k^{\rm{u}}} {\ln \left( {1 + \frac{{\alpha _k^{\rm{u}}g_{k,n}^{\rm{u}}P_k^{\rm{u}}}}{{ {N_0}{N_{\rm{T}}}N_k^{\rm{u}}W}}} \right)} } \right]
- \frac{{B_k^{\rm{u}}\ln 2}}{{{T_{\rm{s}}}W}} \Bigg\} \Bigg) \Bigg\},\label{eps}
\end{align}
where $V_k^{\rm u}\approx N_k^{\rm{u}}$ is applied. As shown in \cite{Gross2015Delay,sun2019optimizing}, this approximation is accurate when the signal-to-noise ratio (SNR) is higher than $10$~dB, which is the usual case in cellular networks. Since $V_k^{\rm u} < N_k^{\rm{u}}$ in all SNR regimes, \eqref{eps} is an upper bound of the approximation on the decoding error probability.

To guarantee the reliability requirement of URLLC, the decoding error probability in \eqref{eps} should not exceed the maximal threshold of the maximum tolerable decoding error probability, i.e.,
\begin{equation}
\bar{\epsilon}_k^{{\rm d,u}} \leq \epsilon^{\max,{\rm u}}. \label{err}
\end{equation}
%In this case, constraint \eqref{err} can be written in a form of $g(\boldsymbol{p},\boldsymbol{\beta}) \leq  \boldsymbol{c}$ as in \eqref{general}, where $\bar{\epsilon}_k^{{\rm d,u}}$ is a function of $P_k^{\rm u}$ and $N_k^{\rm u}$.

\subsection{Problem Formulation}
Improving resource utilization efficiency, such as energy efficiency and spectrum efficiency, is an urgent task in future cellular networks \cite{gui20206g}. In this paper, we take the problem of power minimization as an example to illustrate our method. By changing the objective function, our method can be easily extended to other resource allocation problems.

The total power consumption of a BS consists of the transmit power and the circuit power, given by \cite{Debaillie2015PowerFormu}
\begin{align}
P_{\rm tot} = \frac{1}{\rho} \sum_{ k\in \cal K^{\xi} }P_k^\xi + P^{\rm ca}N_{\rm T} \sum_{ k\in \cal K^{\xi} }N_k^\xi + P_0^{\rm c},\label{total}
\end{align}
where $\rho \in (0,1]$ is the power amplifier efficiency, $P^{\rm ca}$ is the power consumption by each antenna for signal processing on each subcarrier, $P_0^{\rm c}$ is the fixed circuit power consumption.

To save the power consumption of the BS, we minimize $P_{\rm tot}$ subject to QoS constraints, i.e.,
\begin{align}
&\min_{P_k^\xi,N_k^{\xi}} \;
P_{\rm tot},
\label{obj_tot}
\\
\text{s.t.} \;
& \sum_{ k\in \cal K^{\xi} } N_k^{\xi} \leq  N^{\rm max}, \label{sb1_tot} \tag{\theequation a}\\
& \sum_{ k\in \cal K^{\xi} } P_k^{\xi} \leq  P^{\rm max}, \label{sb2_tot} \tag{\theequation b}
\\
& \eqref{dtCon}, \eqref{dsCon},\;\text{and}\; \eqref{err}. \nonumber
\end{align}
where \eqref{sb1_tot} and \eqref{sb2_tot} are the constraints on the total number of subcarriers and the maximum transmit power of the BS. Problem \eqref{obj_tot} is an MINLP problem, which is non-convex. The left-hand sides of constraints \eqref{dtCon}, \eqref{dsCon}, and \eqref{err} do not have closed-form expressions. Thus, finding the global optimal solution is very challenging, especially when the resource allocation should be adjusted according to the dynamic wireless channels and the features of packet arrival processes.

\section{Supervised Deep Learning --- Cascaded Neural Networks for QoS Guarantee}
\label{supervisedLearning}
In this section, we apply supervised deep learning in resource allocation. Specifically, we use two kinds of NNs to approximate the optimal policy that maps the system states to the optimal resource allocation: FNN and cascaded NNs. To obtain labeled training samples, we develop an optimization algorithm to find the global optimal solutions of the problem. Then, we illustrate how to train the cascaded NNs. Finally, we analyze the complexity of the supervised deep learning approaches.

\subsection{FNN for Resource Allocation}
An FNN consists of multiple layers of neurons. Each neuron includes a non-linear activation function and some parameters to be optimized in the training phase \cite{lecun2015deep}. Denote the input and output vectors of the $l$-th layer as $\boldsymbol{x}^{[l]}$ and $\boldsymbol{y}^{[l]}$ respectively. Then, from the activation function and parameters in the $l$-th layer, the output vector can be expressed as follows,
\begin{equation}\label{eq:forward}
\boldsymbol{y}^{[l]} = \delta_{\rm a} (\boldsymbol{W}^{[l]} \boldsymbol{x}^{[l]} + \boldsymbol{b}^{[l]}),
\end{equation}
where $\delta_{\rm a}(\cdot)$ is the activation function, $\varLambda \triangleq \{\boldsymbol{W}^{[l]}, \boldsymbol{b}^{[l]}, l = 0,...,L_{\rm FNN}\}$ are the parameters of the FNN and $L_{\rm FNN}$ is the number of layers. We will use $\text{ReLU}(\cdot)=\max(0, \cdot)$ as the activation function in the rest of this paper unless otherwise specified.

In problem \eqref{obj_tot}, the resource allocation policy depends on the large-scale channel gains, $\boldsymbol{\alpha} = [\alpha_1^\xi,... ,\alpha_K^\xi ]^{\rm T}$, and the packet arrival processes of different kinds of services, where $(\cdot)^{\rm T}$ denotes the transpose operator. More specifically, for delay-tolerant services, the average service rate requirements are determined by the average arrival rates, $[\bar{a}_1, ..., \bar{a}_{|{\mathcal{K}}^{\rm t}|}]$. For delay-sensitive services, the effective capacities of the service processes should be equal to or higher than the effective bandwidth of the arrival processes, $[E_1^{B, \rm s} , ..., E_{|{\mathcal{K}}^{\rm s}|}^{B, \rm s}]$. For URLLC services, the numbers of bits to be transmitted in each TTI depend on the packet sizes of different users, $[ B_1^{\rm u}, ..., B_{|{\mathcal{K}}^{\rm u}|}^{\rm u}] $. The features of the packet arrival processes of all kinds of services are denoted by $\boldsymbol{c}=[{\boldsymbol{c}^{\rm t}},{\boldsymbol{c}^{\rm s}},{\boldsymbol{c}^{\rm u}}]^{\rm T}$, where $\boldsymbol{c}^{\rm t} = [\bar{a}_1,...,\bar{a}_{|{\mathcal{K}}^{\rm t}|}]$, $\boldsymbol{c}^{\rm s} = [E_1^{B, \rm s} , ..., E_{|{\mathcal{K}}^{\rm s}|}^{B, \rm s}]$, and $\boldsymbol{c}^{\rm u} = [ B_1^{\rm u}, ..., B_{|{\mathcal{K}}^{\rm u}|}^{\rm u}] $. The optimal policy of problem \eqref{obj_tot} that maps the features of channels and packet arrival processes to the optimal resource allocation is denoted by $\pi^*$,
\begin{align}\label{eq:opt}
\pi ^* : \boldsymbol{X} \rightarrow \boldsymbol{Y}^*,
\end{align}
where $\boldsymbol{X} = [\boldsymbol{\alpha}^{\rm T} , \boldsymbol{c}^{\rm T}]^{\rm T}$, $\boldsymbol{Y}^* = [{\boldsymbol{P}^*}^{\rm T}, {\boldsymbol{N}^*}^{\rm T}]^{\rm T}$, $\boldsymbol{P}^* = [P_1^{\xi*},... ,P_K^{\xi*} ]^{\rm T}$, and $\boldsymbol{N}^* = [N_1^{\xi*},... ,N_K^{\xi*} ]^{\rm T}$.

As indicated in the universal approximation theorem of NNs \cite{HORNIK1989UniversalApprox}, FNN is a universal approximator of deterministic and continuous functions defined over compact sets. The approximation errors approach to zero as the number of neurons goes to infinite. For our problem, the output of the FNN, denoted by $\tilde{\boldsymbol{Y}}$, includes transmit power and bandwidth allocation, ${\tilde{\boldsymbol{P}}}$ and ${\tilde{\boldsymbol{N}}}$, i.e., $\tilde{\boldsymbol{Y}} \triangleq [{\tilde{\boldsymbol{P}}}^{\rm T}, {\tilde{\boldsymbol{N}}}^{\rm T}] ^{\rm T}$. With this approximation, there are two kinds of errors that will deteriorate the QoS. First, since the number of neurons is finite in the FNN, approximation errors are inevitable, i.e., $\tilde{\boldsymbol{Y}}$ will not be the same as $\boldsymbol{Y}^*$ with probability one. Second, the output of an FNN is continuous, but the numbers of subcarriers are integers. The quantization errors will further deteriorate the QoS of different services.

%
%\blue{However, the approximation over discrete optimization variables can lead to an inaccurate result. This inaccuracy is caused by two types of errors with FNN: the approximation errors during learning and the quantization errors by mapping the continuous output of FNN to discrete values. As a result, the output of the FNN cannot guarantee the QoS.  T, we propose a structure of cascaded neural networks.}

\subsection{Cascaded Neural Networks for QoS Guarantee} \label{sec:cascaded}
\begin{figure}[htbp]
	\vspace{-0.2cm}
	\centering
	\begin{minipage}[t]{0.7\textwidth}
		\centering
		\includegraphics[width=1\textwidth]{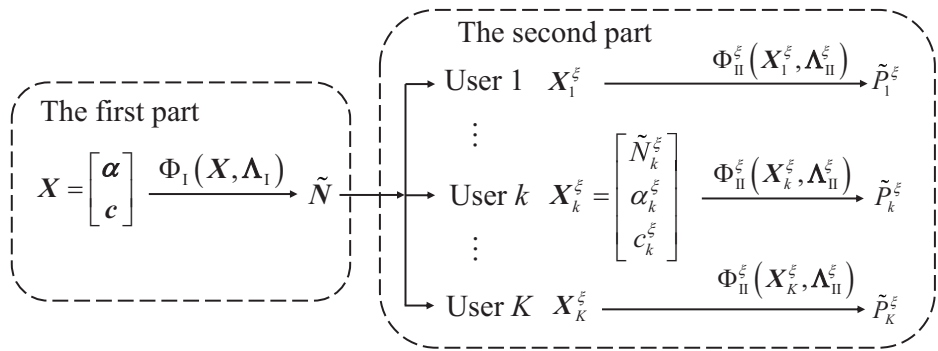}
	\end{minipage}
	\vspace{-0.6cm}
	\caption{Illustration of the cascaded NNs.}
	\label{cas}
	\vspace{-0.2cm}
\end{figure}

To improve the accuracy of the approximation and to ensure the QoS requirements of an MINLP, we propose a cascaded structure consisting of two parts of NNs in Fig. \ref{cas}. The first NN maps the system states to the discrete variables, i.e., $\tilde{\boldsymbol{N}} = \Phi_{\rm I}(\boldsymbol{X},\varLambda_{\rm I})$, where $\varLambda_{\rm I}$ is the parameters of the NN. Like an FNN, the first NN will introduce quantization errors. To alleviate the effect of quantization errors on the QoS, we train another NN that maps the obtained bandwidth allocation to the transmit power that is required to guarantee the QoS constraint of each user. Specifically, in a system with $K$ users, the second part of the cascaded structure consists of $K$ NNs. Each of them approximates the power allocation policy, $\tilde{P}^{\xi}_k = \Phi_{\rm II}^{\xi}(\boldsymbol{X}_k^{\xi},\varLambda_{\rm II}^{\xi})$, where the input of the $k$-th NN is defined as $\boldsymbol{X}_k^{\xi} \triangleq [\tilde{N}^{\xi}_k,\alpha^{\xi}_k,c^{\xi}_k]^{\rm T}$. For the users that request the same type of services, the required transmit power depends on $\alpha^{\xi}_k$ and $c^{\xi}_k$. Since the values of $\alpha^{\xi}_k$ and $c^{\xi}_k$ are included the input of $\Phi_{\rm II}^{\xi}$, we only need to train one NN for all the users that request the same type of service.

Denote the approximation accuracy of the power allocation policy as $\Delta_P$, which is defined as a threshold that satisfies the following requirement,
\begin{align}
\Pr\{ |\Phi_{\rm II}^{\xi}(\boldsymbol{X}_k^{\xi},\varLambda_{\rm II}^{\xi})-P_k^{\xi}(\tilde{N}_k^{\xi} )| \leq \Delta_P\} \geq P_{\rm req}, \label{eq:accurate}
\end{align}
where $P_{\rm req}$ is the required probability with QoS guarantee and $P_k^{\xi}(\tilde{N}_k^{\xi} )$ is the minimum transmit power that is required to satisfy the constraint in \eqref{dtCon}, \eqref{dsCon} or \eqref{err}.\footnote{The minimum transmit power that is required to satisfy the QoS constraints depends on $\alpha^{\xi}_k$ and $c^{\xi}_k$. Since $\alpha^{\xi}_k$ and $c^{\xi}_k$ are two system parameters that do not change with bandwidth allocation, the required transmit power is denoted by $P_k^{\xi}(\tilde{N}_k^{\xi} )$ for notational simplicity.} If the BS allocate $\tilde{P}^{\xi}_k + \Delta_P$ transmit power to the $k$-th user, its' QoS requirement can be satisfied with probability $P_{\rm req}$.

%{By reserving extra transmit power $\Delta_P$ to each user according to $\tilde{P}^{\xi}_k$}, the QoS requirements of each user can be satisfied with probability $P_{\rm req}$. {In the testing phase, the transmit power  is allocated to each user after obtaining the output of cascaded NNs.}

If an FNN is used to approximate the bandwidth and transmit power allocation policy, the quantization errors of $\tilde{\boldsymbol{N}}$ and the approximation errors of $\tilde{\boldsymbol{P}}$ are intertwined. The cascaded NNs can achieve higher accuracy than the FNN due to the following two reasons. First, the quantization errors of $\tilde{\boldsymbol{N}}$ and the approximation errors of $\tilde{\boldsymbol{P}}$ are decoupled. If the approximation of $\Phi_{\rm I}$ is inaccurate, $\tilde{{N}}^{\xi}_k$ will be different from the optimal subcarrier allocation. However, the QoS constraints can be satisfied for any given value of $\tilde{{N}}^{\xi}_k$ if $\Phi_{\rm II}^{\xi}$ outputs the required minimum transmit power. Thus, whether the QoS constraints can be satisfied or not only depends on the approximation accuracy of $\Phi_{\rm II}^{\xi}$ and does not depend on the approximation and quantization errors of $\Phi_{\rm I}$. Second, the dimensions of the input and output of $\Phi_{\rm II}^{\xi}$ are much smaller than that of the FNN. It is much easier to obtain an accurate approximation of the power allocation policy for each user than to obtain an accurate approximation of bandwidth and power allocation policy for all the users. We will validate the performance of them via simulation.

%This will be different with the cascaded NNs. Due to the approximation errors and quantization errors of $\tilde{\boldsymbol{N}}$, the output of $\Phi_{\rm I}$ will be different from the optimal bandwidth allocation. If $\Phi_{\rm II}^{\xi}$ outputs the required minimum transmit power for any given value of $\tilde{{N}}^{\xi}_k$, then the QoS constraint can be satisfied with inaccurate bandwidth allocation. With the cascaded NNs, quantization errors of $\tilde{\boldsymbol{N}}$ and the approximation errors of $\tilde{\boldsymbol{P}}$ are decoupled, and hence the quantization errors and the approximation accuracy of $\Phi_{\rm I}$ have little impact on $\Delta_P$. Therefore, compared with the FNN, the cascaded NNs can guarantee the QoS of different services with a higher probability.

\subsection{Labeled Training Samples}
\label{collection}
The optimal solutions of an MINLP problem can be found by some well-known algorithms, such as branch-and-bound (BnB) \cite{BnB}. However, BnB requires a very high computational complexity, possibly approaching to the exhaustive search for some worst cases \cite{HE2014BnB}.
%\red{In addition, it limits the application only on the problem with differentiable functions, not applicable to that with non-differentiable functions, e.g., \eqref{ru}.} \blue{Does BnB requires the constraints to be differentiable? Why \eqref{ru} is not differentiable?}

To obtain a large number of training samples, we develop an optimization algorithm that converges to the global optimal solution of problem \eqref{obj_tot} with acceptable complexity. First, we validate the feasibility of problem \eqref{obj_tot}, i.e., whether the radio resources, $N^{\rm max}$ and $P^{\max}$, can guarantee the QoS requirements of all the $K$ users. If the problem is feasible, then we find the optimal solution of problem \eqref{obj_tot}.

\subsubsection{Feasibility of Problem \eqref{obj_tot}} To find out whether problem \eqref{obj_tot} is feasible, we minimize the required total transmit power $\sum_{ k\in \cal K^{\xi} }P_k^\xi$ subject to the other constraints, if the required total transmit power is less than $P^{\max}$, then the problem is feasible. Otherwise, it is infeasible. The required minimum total transmit power can be found by solving the following problem,
\begin{align}
&\min_{P_k^\xi,N_k^{\xi}} \;
\sum_{ k\in \cal K^{\xi} }P_k^\xi,
\label{obj}
\\
\text{s.t.} \;
%& \sum_{ k\in \cal K^{\xi} } N_k^{\xi} \leq  N^{\rm max}, \label{sb1} \tag{\theequation a}\\
%& \sum_{ k\in \cal K^{\xi} } P_k^{\xi} \leq  P^{\rm max}, \label{sb2} \tag{\theequation b}
%\\
& \eqref{sb1_tot}, \eqref{dtCon}, \eqref{dsCon},\;\text{and}\; \eqref{err}, \nonumber
\end{align}

\label{algo}
To solve problem \eqref{obj}, we find the optimal bandwidth allocation that minimizes the total required transmit power. For a given bandwidth allocation, $N_k^\xi$, the expressions in \eqref{rb}, \eqref{EC2} and \eqref{eps} are monotonous with respect to $P_k^\xi$. If we can drive the closed-form expressions of the multiple integrals in \eqref{rb}, \eqref{EC2} and \eqref{eps}, then the binary search can be used to obtain the minimum transmit power that is required to ensure constraints \eqref{dtCon}, \eqref{dsCon}, and \eqref{err}, $P_k^\xi(N_k^\xi)$. However, the closed-form expressions are not available, and this approach is time-consuming since the system needs to compute the multiple integrals in each iteration of the binary search. To avoid computing the integrals, we adopt the stochastic gradient descent (SGD) method to find the minimum transmit power subject to constraints \eqref{dtCon}, \eqref{dsCon}, and \eqref{err}, respectively. As shown in \cite{lee2019deep,sun2019GC}, the SGD method is efficient in solving constrained optimization problems, where some constraints do not have closed-form expressions.

Let $x(\tau)$ be a variable obtained in the $\tau$-th iteration. For delay-tolerant services, by substituting \eqref{rb} into \eqref{dtCon}, the optimal transmit power with a given bandwidth allocation can be found through the following iterations,
\begin{equation}\label{eq:SGD1}
P_k^{\rm t} (\tau+1) = \left[
P_k^{\rm t} (\tau) + \phi (\tau) \left(
{\bar{a}_k} - \bar{R}_k^{{\rm{t}}}(\tau)
 \right)
 \right] ^+ ,
\end{equation}
where $[x]^+ = \max\{x, 0\}$, $\phi (\tau)>0$ is the step size, and $\bar{R}_k^{{\rm{t}}}(\tau)$ is the average service rate in \eqref{rb}, which is estimated from a set of realizations of small-scale channel gains on the $N_k^{\rm t}$ subcarriers. From \eqref{eq:SGD1}, we can see that if $\bar{R}_k^{{\rm{t}}}(\tau) > \bar{a}_k$, then $P_k^{\rm t} (\tau+1) > P_k^{\rm t} (\tau)$. Otherwise, $P_k^{\rm t} (\tau+1) < P_k^{\rm t} (\tau)$. As indicated in \cite{bottou-98x}, with $\phi (\tau) \sim \mathcal{O}(1/\tau)$, the SGD method can converge to the unique optimal transmit power that satisfies $\bar{R}_k^{{\rm{t}}} = {\bar{a}_k}$.

To obtain an unbiased gradient estimation with the SGD method, the expectation in the constraint should  be linear \cite{sun2019GC}. For delay-sensitive services, we first transform constraint \eqref{dsCon} into an equivalent form that is linear to the expectation, i.e., $\mathbb{E}_{g_{k,n}^{\rm s}} \left[ \exp \left(
- \theta_k^{\rm s}{T_{\rm c}} R_k^{\rm s}
\right)\right]
-
\exp(-\theta_k^{\rm s}{T_{\rm c}} E_k^{B, \rm s} )
\leq 0$. Then, the optimal transmit power with a given bandwidth allocation can be found through the following iterations,
\begin{equation}\label{eq:SGD2}
P_k^{\rm s} (\tau+1) = \left[
P_k^{\rm s} (\tau) + \phi (\tau) \left(
\exp(- \theta_k^{\rm s}{T_{\rm c}} R_k^{\rm s}(\tau))
- \exp(-\theta_k^{\rm s}{T_{\rm c}} E_k^{B, \rm s} )
\right)
\right] ^+,
\end{equation}
where $R_k^{\rm s}(\tau)$ is the realization of the achievable rate in \eqref{Rs}.

For URLLC services, the optimal transmit power with a given bandwidth allocation can be obtained from the following iterations,
\begin{align}
P_k^{\rm u} (\tau+1)  =
\left[
P_k^{\rm u} (\tau) + \phi (\tau) \left(
\bar{\epsilon}_k^{\rm d,u}(\tau) - \epsilon^{\max, \rm u}
\right)
\right] ^+ ,\label{eq:SGD3}
\end{align}
where $\bar{\epsilon}_k^{\rm d,u}(\tau)$ is the realization of the decoding error probability in \eqref{eps}.

\begin{table}[hbt]
	\caption{Bandwidth Allocation Algorithm for Solving Problem \eqref{obj}}
	\vspace{-1.0cm}
	\begin{tabular}{p{17cm}}
		\\\hline
	\end{tabular}
	\label{sub}
	\begin{algorithmic}[1]
    \vspace{-0.5cm}
		\REQUIRE {Large-scale channel gains $\alpha_k^\xi$ and QoS constraints $c_k^\xi$.}
		\STATE Initialize $N_k^{\xi} = 1, \forall k \in \cal K^{\xi}$.
		\STATE Compute $\Delta P_k^{\xi}(N_k^{\xi}) = P_k^{\xi}(N_k^{\xi} ) - P_k^{\xi}({N}_k^{\xi}+1), \forall k \in \cal K^{\xi}$.
		\WHILE{ $\sum_{ k\in \cal K^{\xi} } N_k^{\xi} \leq  N^{\rm max}$ and $\Delta P_{k^*}^{\xi}(N_{k^*}^{\xi})>0$}		
		\STATE $k^* := \arg\max_{k \in \cal K^{\xi}} \Delta P_k^{\xi} (N_k^{\xi})$.
		\STATE $N_{k^*}^{\xi} := N_{k^*}^{\xi} + 1$.
        \STATE Update $P_{k^*}^{\xi}(N_{k^*}^{\xi} ) $ and $ P_{k^*}^{\xi}({N}_{k^*}^{\xi}+1)$ according to \eqref{dtCon}, \eqref{dsCon}, and \eqref{err}.
		\STATE $\Delta P_{k^*}^{\xi}(N_{k^*}^{\xi}) := P_{k^*}^{\xi}(N_{k^*}^{\xi} ) - P_{k^*}^{\xi}({N}_{k^*}^{\xi}+1)$.
%		\IF {$\sum_{ k\in \cal K^{\xi} } N_k^{\xi} =  N^{\rm max}$}
%		\STATE Break.%\COMMENT{\textbf{Add in Algorithm \ref{sub2}}}
%		\ENDIF
		\ENDWHILE
		\RETURN {$ \hat{N}_k^{\xi}:= N_{k}^{\xi}$ and $\hat{P}_k^{\xi}(\hat{N}_{k}^{\xi} ):=P_{k}^{\xi}(N_{k}^{\xi} )$, $k=1,...,K.$ }
\vspace{-0.3cm}
	\end{algorithmic}
	\vspace{-0.3cm}
	\begin{tabular}{p{17cm}}
		\\\hline
	\end{tabular} \vspace{-0.5cm}
\end{table}

The bandwidth allocation algorithm for solving problem \eqref{obj} is shown in Table \ref{sub}.

\emph{Step 1:} Initialize bandwidth allocation with $N_k^{\xi} = 1, \forall k$, and compute $\Delta P_k^{\xi}(N_k^{\xi}) = P_k^{\xi}(N_k^{\xi} ) - P_k^{\xi}({N}_k^{\xi}+1), \forall k \in \cal K^{\xi}$ with the SGD method.

\emph{Step 2:} Assign one more subcarrier to the user with the highest power saving, i.e.,
\begin{align}
k^* = \arg\max_{k \in \cal K^{\xi}} \Delta P_k^{\xi} (N_k^{\xi}).\nonumber
\end{align}

\emph{Step 3:} Update $N_{k^*}^{\xi}$ and $\Delta P_{k^*}^{\xi}(N_{k^*}^{\xi})$ with the SGD method.

Finally, we execute Step 2 and Step 3 iteratively until $\sum_{ k\in \cal K^{\xi} } N_k^{\xi} =  N^{\rm max}$.

\subsubsection{Algorithm for Solving Problem \eqref{obj_tot}}
\label{algo2} If problem \eqref{obj_tot} is feasible, we use the algorithm in Table \ref{sub2} to solve this problem.

\emph{Step 1 (Lines 2-10 in Table \ref{sub2}):} We find the optimal bandwidth and transmit power allocation that minimizes $P_{\rm tot}$ without the total transmit power constraint in \eqref{sb2_tot}. To achieve this goal, we replace $\Delta P_k^{\xi}(N_k^{\xi})$ in Table~\ref{sub} with
\begin{align}
\Delta P_{\mathrm{tot}, k}^{\xi}(N_k^{\xi}) & \triangleq P_{\rm tot}([N_1^{\xi}, ...,N_k^{\xi}, ..., N_K^{\xi}])-P_{\rm tot}([N_1^{\xi}, ..., N_k^{\xi}+1, ..., N_K^{\xi}]) \nonumber\\
&= \Delta P_k^{\xi}(N_k^{\xi}) - P^{\rm ca}N_{\rm T}.\label{eq:deltaPtot}
\end{align}
Like the algorithm in Table \ref{sub}, each subcarrier is allocated to the user with the highest $\Delta P_{\mathrm{tot}, k}^{\xi}(N_k^{\xi})$. The results obtained in this step is denoted by $ \check{N}_k^{\xi}$ and $\check{P}_k^{\xi}(\check{N}_{k}^{\xi} )$. If the equality in constraint \eqref{sb1_tot} holds with the results in this step, i.e., $\sum_{ k\in \cal K^{\xi} } \check{N}_k^{\xi} =  N^{\rm max}$, the solutions obtained from the algorithms in Tables \ref{sub} and \ref{sub2} are the same. This is because the second term in \eqref{total} is fixed, and thus minimizing $\sum_{ k\in \cal K^{\xi} }P_k^\xi$ is equivalent to minimizing $P_{\rm tot}$.

\emph{Step 2 (Lines 11-21 in Table \ref{sub2}):} If $\sum_{ k\in \cal K^{\xi} } \check{N}_k^{\xi} <  N^{\rm max}$, then we check whether constraint \eqref{sb2_tot} is satisfied or not. If it is satisfied, $\check{N}_k^{\xi}$ and $\check{P}_k^{\xi}(\check{N}_{k}^{\xi} )$ will be returned as the outputs of the algorithm. Otherwise, more subcarriers will be assigned to the users until the transmit power constraint is satisfied (Lines 13-21 in Table \ref{sub2}).

% based on the results of $\check{N}_k^{\xi}$ and $\check{P}_k^{\xi}(\check{N}_{k}^{\xi} )$ (we reuse the results from Line 9 to avoid repetitive calculations) to decrease the transmit power until constraint \eqref{sb2_tot} is satisfied, as in Line 18-20. To this end, the algorithm in Table \ref{sub2} can find the optimal resource allocation scheme for problem \eqref{obj_tot}.

%\blue{The computational complexity of the algorithm for solving problem \eqref{obj_tot} lies on two parts of computation, i.e., $\Delta P_k^\xi(N_k^\xi)$ and $\Delta P_{\mathrm{tot}, k}^{\xi}(N_k^{\xi})$. The first part of the computation is the feasibility assurance by finding the minimum total transmit power $\sum_{ k\in \cal K^{\xi} }P_k^\xi$ as in Line 1 of Table \ref{sub2}. That requires $N^{\max}$ updates on $\Delta P_k^\xi(N_k^\xi)$ as discussed in Section \ref{algo}. The second part is the computation for finding the minimized $P_{\rm tot}$ (Line 4-9 and Line 13-21 of Table \ref{sub2}). As we reuse the results from Line 4-9 if constraint \eqref{sb2_tot} is not satisfied, the maximum number of updates on $\Delta P_{\mathrm{tot}, k}^{\xi}(N_k^{\xi})$ is $N^{\max}$. As a result, taking account of these two parts of computations, the complexity of the algorithm for solving problem \eqref{obj_tot} linearly increases with $2N^{\max}$.}

\begin{table}[hbt]
	\caption{Bandwidth Allocation Algorithm for Solving Problem \eqref{obj_tot}}
	\vspace{-1.0cm}
	\begin{tabular}{p{17cm}}%{p{8.5cm}}
		\\\hline
	\end{tabular}
	\label{sub2}
	\begin{algorithmic}[1]
\vspace{-0.5cm}
		\REQUIRE {Large-scale channel gains $\alpha_k^\xi$ and QoS constraints $c_k^\xi$.}
%		\\\hrulefill
		\STATE Check whether problem \eqref{obj_tot} is feasible or not with the algorithm in Table \ref{sub}.
		%		\\\hrulefill
%		\STATE Replace $\Delta P_k^{\xi}(N_k^{\xi})$ in Algorithm \ref{sub} with $P_{\rm tot}^{\xi}([N_1^{\xi}, ...,N_k^{\xi}, ..., N_K^{\xi}])-P_{\rm tot}^{\xi}([N_1^{\xi}, ..., N_k^{\xi}+1, ..., N_K^{\xi}])$ and remove the while condition of $\sum_{ k\in \cal K^{\xi} } N_k^{\xi} \leq  N^{\rm max}$ in Line 3 of Algorithm \ref{sub}.
		%
		\STATE Initialize $N_k^{\xi} := 1, \forall k \in \cal K^{\xi}$, and $P_{\rm tot}([1, ...,1])$.
		\STATE Compute $\Delta P_{\mathrm{tot}, k}^{\xi}(N_k^{\xi}) := P_{\rm tot}([N_1^{\xi}, ...,N_k^{\xi}, ..., N_K^{\xi}])-P_{\rm tot}([N_1^{\xi}, ..., N_k^{\xi}+1, ..., N_K^{\xi}]), \forall k \in \cal K^{\xi}$.
		\WHILE{ $\sum_{ k\in \cal K^{\xi} } N_k^{\xi} \leq  N^{\rm max}$ and $\Delta P_{\mathrm{tot}, k^*}^{\xi}(N_{k^*}^{\xi})>0$}		
		\STATE $k^* := \arg\max_{k \in \cal K^{\xi}} \Delta P_{\mathrm{tot},k}^{\xi} (N_k^{\xi})$.
		\STATE $N_{k^*}^{\xi} := N_{k^*}^{\xi} + 1$.
        \STATE Update $P_{\rm tot}([N_1^{\xi}, ...,N_{k^*}^{\xi}, ..., N_K^{\xi}])$ and $P_{\rm tot}([N_1^{\xi}, ..., N_{k^*}^{\xi}+1, ..., N_K^{\xi}])$ according to \eqref{dtCon}, \eqref{dsCon}, \eqref{err}, and \eqref{total}.
		\STATE $\Delta P_{\mathrm{tot}, k^*}^{\xi}(N_{k^*}^{\xi}) :=
		P_{\rm tot}([N_1^{\xi}, ...,N_{k^*}^{\xi}, ..., N_K^{\xi}])-P_{\rm tot}([N_1^{\xi}, ..., N_{k^*}^{\xi}+1, ..., N_K^{\xi}])$.
		\ENDWHILE
		\STATE {$ \check{N}_k^{\xi} := N_{k}^{\xi}$ and $\check{P}_k^{\xi}(\check{N}_{k}^{\xi} ):=P_{k}^{\xi}(N_{k}^{\xi} )$, $\forall k=1,...,K$.}
		\IF {$\sum_{ k\in \cal K^{\xi} } \check{P}_k^{\xi}(\check{N}_k^{\xi}) \leq  P^{\rm max}$}
		\RETURN {$ \dot{N}_k^{\xi} := \check{N}_k^{\xi}, \dot{P}_k^{\xi}(\dot{N}_{k}^{\xi}) := \check{P}_k^{\xi}(\check{N}_{k}^{\xi})$. }
		\ELSE
		\STATE {$ \dot{N}_k^{\xi} := \check{N}_k^{\xi}, \dot{P}_k^{\xi}(\dot{N}_{k}^{\xi}) := \check{P}_k^{\xi}(\check{N}_{k}^{\xi})$. }
		\WHILE{$\sum_{ k\in \cal K^{\xi} } \dot{P}_k^{\xi}(\dot{N}_k^{\xi}) \leq  P^{\rm max}$}		
		\STATE $k^* := \arg\max_{k \in \cal K^{\xi}} \Delta \dot{P}_k^{\xi} (\dot{N}_k^{\xi})$.
		\STATE $\dot{N}_{k^*}^{\xi} := \dot{N}_{k^*}^{\xi} + 1$.
        \STATE Update $\dot{P}_{k^*}^{\xi}(\dot{N}_{k^*}^{\xi} ) $ and $ \dot{P}_{k^*}^{\xi}(\dot{N}_{k^*}^{\xi}+1)$ according to \eqref{dtCon}, \eqref{dsCon}, and \eqref{err}.
		\STATE $\Delta \dot{P}_{k^*}^{\xi}(N_{k^*}^{\xi}) := \dot{P}_{k^*}^{\xi}(\dot{N}_{k^*}^{\xi} ) - \dot{P}_{k^*}^{\xi}(\dot{N}_{k^*}^{\xi}+1)$.
		\ENDWHILE
		\RETURN {$ \dot{N}_k^{\xi}, \dot{P}_k^{\xi}(\dot{N}_{k}^{\xi} )$.}
		\ENDIF
\vspace{-0.3cm}
	\end{algorithmic}
	\vspace{-0.3cm}
	\begin{tabular}{p{17cm}}%{p{8.5cm}}
		\\\hline
	\end{tabular} \vspace{-0.2cm}
\end{table}

\subsubsection{Optimality of Algorithm for Solving Problem \eqref{obj_tot}}
In this subsection, we first discuss the optimality conditions of the algorithm in Table \ref{sub}, and prove that the conditions are satisfied with all the three kinds of services. Then, we prove the optimality of the algorithm in Table \ref{sub2}. %Based on the optimality discussion for the algorithm in Table \ref{sub}, we extend its optimality to the algorithm in Table \ref{sub2} for problem \eqref{obj_tot}.

The algorithm in Table \ref{sub} can find the global optimal solution for problem \eqref{obj} if the following two conditions hold (See proof in Appendix \ref{App:Prop1}).

\begin{condition}\label{Condition1}
	\emph{$P_k^{\xi}(N_k^{\xi}) > P_k^{\xi}(N_k^{\xi}+1)$, $\forall N_k^{\xi} = 1, ..., N^{\rm max}-1.$}
\end{condition}
Condition 1 means the required transmit power decreases with the number of subcarriers.

\begin{condition}\label{Condition2}
	\emph{$\Delta P_k^{\xi}(N_k^{\xi}) \geq \Delta P_k^{\xi}(N_k^{\xi}+1), \forall N_k^{\xi} = 1, ..., N^{\rm max}-1.$ }
\end{condition}

For notational simplicity, we denote the left-hand sides of constraints \eqref{dtCon}, \eqref{dsCon}, and \eqref{err} by $f_k^\xi( P_k^\xi, N_k^\xi )$, $\xi \in \{\rm{t},\rm{s},\rm{u}\}$, respectively. Since the minimum transmit power is obtained when the equalities in these constraints hold, we can prove the following proposition (see proof in Appendix \ref{App:Prop2}).

%\blue{For condition \ref{Condition2}, it can be satisfied if $P_k^{\xi}(N_k^{\xi})$ is convex. (Do we need this sentence? It interrupts the logic between the previous sentence and the property.)}

%if there is a continuous relaxation of the constraint, f(P,N)=C, where N \in R^+, and f(P,N) is jointly concave in P,N

\begin{prop}
	\label{P2}
	For a constraint $f_k^\xi( P_k^\xi, N_k^\xi ) = c^{\xi}_k, P_k^\xi \in \mathbb{R}^+, N_k^\xi\in \mathbb{Z}^+$, if there exists a continuous relaxation of the constraint,  ${\mathord{\buildrel{\lower3pt\hbox{$\scriptscriptstyle\smile$}}\over f} }_k^\xi( P_k^\xi, {\mathord{\buildrel{\lower3pt\hbox{$\scriptscriptstyle\smile$}}\over N} }_k^\xi ) = c^{\xi}_k$, where ${\mathord{\buildrel{\lower3pt\hbox{$\scriptscriptstyle\smile$}}\over N} }_k^\xi\in \mathbb{R}^+$ and $\mathord{\buildrel{\lower3pt\hbox{$\scriptscriptstyle\smile$}} \over f} _k^\xi(P_k^\xi,\mathord{\buildrel{\lower3pt\hbox{$\scriptscriptstyle\smile$}} 	\over N} _k^\xi )$ is jointly concave (or convex) in $P_k^\xi$, ${\mathord{\buildrel{\lower3pt\hbox{$\scriptscriptstyle\smile$}}\over N} }_k^\xi$ and increases (or decreases) with $P_k^\xi$ and ${\mathord{\buildrel{\lower3pt\hbox{$\scriptscriptstyle\smile$}}\over N} }_k^\xi$, then Conditions \ref{Condition1} and \ref{Condition2} hold for the original constraint, $f_k^\xi( P_k^\xi, N_k^\xi ) = c^{\xi}_k, P_k^\xi \in \mathbb{R}^+, N_k^\xi\in \mathbb{Z}^+$.
\end{prop}

\emph{Delay-Tolerant Services:} As proved in \cite{Zhikun2013Energy}, \eqref{rb} is strictly concave in $P_k^{\rm t}$. If $f(x)$ is concave, then $yf(x/y)$ is jointly concave in $x$ and $y$ \cite{boyd}. Thus, \eqref{rb} is jointly concave in $P_k^{\rm t}$ and $N_k^{\rm t}$. In addition, the Shannon's capacity increases with transmit power and the number of subcarriers. Therefore, Condition \ref{Condition1} and \ref{Condition2} hold for delay-tolerant services.

\emph{Delay-Sensitive Services:} According to the results in \cite{EEECLY}, we know that effective capacity is jointly concave in $P_k^{\rm s}$ and $N_k^{\rm s}$ and increases with $P_k^{\rm s}$ and $N_k^{\rm s}$. Therefore, Conditions \ref{Condition1} and \ref{Condition2} also hold for delay-tolerant services.

\emph{URLLC Services:} Unlike the above two types of services, constraint \eqref{err} for URLLC is not convex in $P_k^{\rm u}$ and $N_k^{\rm u} $ in general. To study whether the proposed algorithm can find the optimal solution, we first consider an asymptotic scenario: $N_{\rm T}$ is large. When $N_{\rm T}$ is sufficiently large, due to channel hardening, we have \cite{Rusek2013Scaling}
\begin{align}\label{eq:asym1}
{\ln \left( {1 + \frac{{\alpha _k^{\rm{u}}g_{k,n}^{\rm{u}}P_k^{\rm{u}}}}{{ {N_0}{N_{\rm{T}}}N_k^{\rm{u}}W}}} \right)} \to {\ln \left( {1 + \frac{{\alpha _k^{\rm{u}}P_k^{\rm{u}}}}{{ {N_0}N_k^{\rm{u}}W}}} \right)}.
\end{align}
Then, the minimum transmit power that can satisfy constraint \eqref{err} can be derived as follows,
\begin{align}
P_k^{\rm{u}}(N_k^{\rm{u}}) =
\frac{{{N_0}N_k^{\rm{u}}W}}{{\alpha _k^{\rm{u}}}}\left\{ {\exp \left[ {\frac{{B_k^{\rm{u}}\ln 2}}{{{T_{\rm{s}}}N_k^{\rm{u}}W}} + \frac{{f_Q^{ - 1}\left( {{{\bar \varepsilon }^{\max {\rm{,u}}}}} \right)}}{{\sqrt {{T_{\rm{s}}}N_k^{\rm{u}}W} }}} \right] - 1} \right\}\label{PU}.
\end{align}
According to the analysis in \cite{sun2019optimizing}, $P_k^{\rm{u}}(N_k^{\rm{u}})$ first decreases with $N_k^{\rm{u}}$ and then increases with $P_k^{\rm{u}}(N_k^{\rm{u}})$. We denote $\acute{N}_k^{\rm{u}}$ as the optimal number of subcarriers that minimizes $P_k^{\rm{u}}(N_k^{\rm{u}})$. Since $\Delta P_k^{\rm{u}}(\acute{N}_k^{\rm{u}}) < 0 $, the number of subcarriers assigned to the $k$-th URLLC user will not exceed $\acute{N}_k^{\rm{u}}$. Moreover, $P_k^{\rm{u}}(N_k^{\rm{u}})$ is convex and decreases with $N_k^{\rm{u}}$ in the region $[1,\acute{N}_k^{\rm{u}}]$ \cite{sun2019optimizing}. Therefore, Conditions 1 and 2 hold for URLLC services in the asymptotic scenario.

For non-asymptotic scenarios, the expectation in \eqref{eps} is a $N_k^u$-fold integral. Since the number of folds increases with the optimization variable, $N_k^u$, one can neither derive a closed-form expression nor get any strict proof. When the number of antennas is large, e.g., $N_{\rm T} > 32$, \eqref{PU} is a good approximation of the required transmit power in the non-asymptotic scenarios, and hence Conditions 1 and 2 hold. For systems with small numbers of antennas, we will validate Conditions 1 and 2 via numerical results with typical parameters in 5G cellular networks.

The above analysis indicates that the algorithm in Table \ref{sub} can find the optimal solution of problem \eqref{obj}. In addition, by solving problem \eqref{obj}, we know whether problem \eqref{obj_tot} is feasible or not. If the problem is feasible, the following proposition shows that the algorithm in Table \ref{sub2} can find the optimal solution to the problem.

\begin{prop}
	\label{P3}
	The algorithm in Table \ref{sub2} can find the global optimal solution for problem \eqref{obj_tot} if Conditions \ref{Condition1} and \ref{Condition2} hold.
	\begin{proof}
		See proof in Appendix \ref{App:Prop3}.
	\end{proof}
\end{prop}

\subsection{Train the Cascaded NNs}
\label{dnn_train}

With the algorithm in Table \ref{sub2}, we can obtain a labeled training sample, $\boldsymbol{N}^*$ and $\boldsymbol{P}^*$, for any given input $\boldsymbol{X}$. To obtain enough labeled training samples, we randomly generate a large number of inputs and find the corresponding optimal solutions. One part of the data is used to train the NNs, and the other part of the data is used to test the performance of the NNs.

The parameters of the NNs are initialized with Gaussian distributed random variables with zero mean and unit variance. In each training epoch, a batch of training samples is randomly selected from all the training samples to train the NNs. The parameters of $\Phi_{\rm I}$ are optimized with the Adam algorithm \cite{kingma2014adam} to minimize a loss function, defined as $\mathcal{L}_{\rm I}(\varLambda_{\rm I}) = \frac{1}{M_t} \sum_{m_t = 1}^{M_t} ( \log(\boldsymbol{N}^*_{m_t} +1) - \log(\tilde{\boldsymbol{N}}_{m_t}+1))^2$, where ${M_t}$ is the number of training samples in each batch. Similarly, we optimize the parameters of $\Phi^{\xi}_{\rm I}$ to minimize $\mathcal{L}_{\rm II}(\varLambda_{\rm II}^{\xi}) = \frac{1}{M_t} \sum_{m_t = 1}^{M_t} ( \log(\boldsymbol{P}^*_{m_t} +1) - \log(\tilde{\boldsymbol{P}}_{m_t}+1))^2$. When the value of a loss function is below a required threshold, the difference between the outputs of the NNs and the optimal resource allocation is small enough, and the outputs of the NNs are near-optimal.

%\boldsymbol{y}^{[L]}

%https://peltarion.com/knowledge-center/documentation/modeling-view/build-an-ai-model/loss-functions/mean-squared-logarithmic-error
%When the value of $\mathcal{L}(\varLambda)$ is below a required threshold, $\sigma_{\mathcal{L}}$, the training phase is finished.

\subsection{Complexity of Deep Learning Algorithms}
Since the cascaded NNs are made up of multiple FNNs, we first analyze the complexity of the FNN and then extend the results to the cascaded NNs.

\subsubsection{Fast Resource Allocation} After the training phase, the forward propagation algorithm is applied to compute the output of a neural network for fast resource allocation. The processing time of the forward propagation algorithm is determined by the numbers of three kinds of operations to be executed, i.e., ``$+$", ``$\times$", and ``$\max(0,\cdot)$" in the ReLU function. We first derive the number of multiplications that is required to compute the output of the FNN, which is denoted by $N_{\rm FNN}^{\rm FP}$. According to \eqref{eq:forward}, the numbers of multiplications for computing the output of the $l$-th layer are $n_{\rm FNN}^{[l]}\times n_{\rm FNN}^{[l+1]}$, where $n_{\rm FNN}^{[l]}$ is the number of neurons in the $l$-th layer. Thus, we have
\begin{align}\label{eq:multiFNN}
N_{\rm FNN}^{\rm FP} = \sum_{l=0}^{L_{\rm FNN}-1}n_{\rm FNN}^{[l]}\times n_{\rm FNN}^{[l+1]}.
\end{align}
Since the cascaded NNs consist of multiple FNNs, from \eqref{eq:multiFNN}, we can obtain the numbers of multiplications required to compute the output of the cascaded NNs,
\begin{align}\label{eq:additiveCAS}
N_{\rm CAS}^{\rm FP} = \sum_{l=0}^{L_{\rm I}-1}n_{\rm I}^{[l]}\times n_{\rm I}^{[l+1]} + M_T\sum_{l=0}^{L_{\rm II}-1}n_{\rm II}^{[l]}\times n_{\rm II}^{[l+1]},
\end{align}
where $L_{\rm I}$ and $L_{\rm II}$ are the number of layers of $\Phi_{\rm I}$ and $\Phi_{\rm II}$, respectively, $n_{\rm I}^{[l]}$ and $n_{\rm II}^{[l]}$ are the number of neurons in the $l$-th layer of $\Phi_{\rm I}$ and $\Phi_{\rm II}$, respectively, and $M_T$ is the types of services. It is not hard to see that the number of additive operations is also $N_{\rm CAS}^{\rm FP}$ and the number of ReLU operations is much smaller than $N_{\rm CAS}^{\rm FP}$. Thus, the complexity of the forward propagation algorithm with the cascaded NNs is $\mathcal{O}(N_{\rm CAS}^{\rm FP})$, which is low enough to be implemented in real-world networks for optimizing resource allocation in real time \cite{she2020deep}.

\subsubsection{Training Algorithm} In the training phase, the Adam algorithm is used to optimize the parameters of FNNs, where the backward propagation algorithm is used to compute the gradients of the loss function with respect to the parameters in each layer \cite{kingma2014adam}. Similar to the forward propagation algorithm that computes the output from the first layer to the last layer, the backward propagation algorithm computes the gradient of the loss function with respect to the parameters from the last layer to the first layer. The complexity of the backward propagation algorithm is the same as the forward propagation algorithm, $\mathcal{O}(N_{\rm CAS}^{\rm FP})$. If there are $N_{\rm ep}$ epochs in the training phase and $M_t$ training samples are selected to train the NNs in each epoch, then the computing complexity for training the FNN and the cascaded NNs is $\mathcal{O}(N_{\rm ep}M_t N_{\rm FNN}^{\rm FP})$ and $\mathcal{O} (N_{\rm ep}M_t N_{\rm CAS}^{\rm FP})$, respectively.

\subsubsection{Finding Labeled Training Samples} Before training the NNs, the optimization algorithm in Table \ref{sub2} is applied to find the labeled training samples. In each iteration, the algorithm assigns one more subcarrier to one of the users. Thus, the number of iterations does not exceed $N^{\max}$. Within each iteration, the algorithm needs to compute the value of $\Delta P_{\mathrm{tot}, k}^{\xi}(N_k^{\xi})$ by using the SGD method in \eqref{eq:SGD1}, \eqref{eq:SGD2} and \eqref{eq:SGD3}. Denote the computing complexity of the SGD method by $\Omega_P$. Then, the complexity of the algorithm in Table \ref{sub2} is ${\mathcal{O}}(M_t^{\rm tot}N^{\max}\Omega_P)$, where $M_t^{\rm tot}$ is the total number of labeled training samples. The SGD method is an iterative algorithm and the convergence speed depends on how fast the learning rate decreases and the required accuracy of the final results. As suggested by \cite{bottou-98x}, the learning rate cannot decrease too fast to ensure convergence. In general, it takes thousands of steps to converge to an accurate result, i.e., the transmit power required to guarantee the QoS constraints. Therefore, the optimization algorithm in Table \ref{sub2} can hardly find the optimal resource allocation every few seconds according to the variations of the large-scale channel gains and the features of packet arrival processes.

%, the computing complexity is determined by this algorithm.
%
%Thus, the complexity of the algorithm is ${\mathcal{O}}(N^{\max}\Omega_P)$, where $\Omega_P$ denotes the operations needed to compute $\Delta P_{k^*}^{\xi}(N_{k^*}^{\xi})$. In other words, the complexity of the algorithm linearly increases with $N^{\max}$ and does not change with the total number of users. Since the second term in \eqref{eq:deltaPtot} is constant, to compute the value of , we only need to compute $\Delta P_k^{\xi}(N_k^{\xi})$. Therefore, the complexity of the algorithm in Table \ref{sub2} is the same as that of the algorithm in Table \ref{sub}.

\section{Deep Transfer Learning in Non-Stationary Wireless Networks}
\label{tf}
%In this section, we study how to approximate the optimal policy with a NN in non-stationary wireless networks.

Since the cascaded NNs are trained offline, it only works well in stationary wireless networks. However, real-world wireless networks are highly dynamic and non-stationary. There are a lot of hidden variables that are not included in the input of the cascaded NNs but have significant impacts on the optimal solution. For example, the optimal resource allocation for delay-sensitive and URLLC services depends on distributions of small-scale channel gains as well as the types of services in the network. If these distributions and parameters change, a NN trained offline is no longer a good approximation of the optimal resource allocation policy in the new scenario \cite{riley2019three}. Such an issue is known as the task mismatch problem \cite{Pan2010TFsurvey}.

A straightforward approach is to train a new NN from scratch in a new scenario. When the hidden variables change, the system can hardly obtain a large number of training samples in the new scenario. This is because the algorithm in Table \ref{sub2} cannot be executed in real time\footnote{To execute Lines 7 and 18 of the algorithm in Table \ref{sub2}, the system needs to compute $\Delta P_{k^*}^{\xi}(N_{k^*}^{\xi})$ with an iterative algorithm.}.  To update the NN with a few training samples, we apply deep transfer learning.

\subsection{Preliminary of Deep Transfer Learning}
The learning process is to accomplish a learning \textit{task} based on a data \textit{domain}. According to the definitions in \cite{Pan2010TFsurvey}, a \textit{domain} consists of a feature space and the corresponding marginal probability distribution, e.g., $\boldsymbol{X}$ and its distribution.
A \textit{task} consists of a label space and an objective predictive function that maps from $\boldsymbol{X}$ to $\boldsymbol{Y}^*$. The function is not observed but learned from the training samples, i.e., $\{ \boldsymbol{X}, \boldsymbol{Y}^* \}$. The basic idea of transfer learning is to exploit the knowledge from a well-trained source task to a new target task \cite{Chuanqi2018transfer}.

Fine-tuning is the most widely used method in deep transfer learning \cite{shen2019transfer}. The basic idea is to fix the parameters in the first a few layers and update the parameters in the last a few layers. In deep transfer learning, parts of the well-trained NN of the source task are reused in the NN of the target task. In this way, the number of labeled training samples are needed to fine-tune the new NN is much less than that needed to train a new NN with randomly initialized parameters (i.e., learning from scratch).

\begin{figure}[tbp]  %%
\vspace{-0.2cm}
\centering
\subfigure[{Non-stationary wireless channels.}]{
\label{tf_one2one} %% label for second subfigure
\includegraphics[width=0.35\textwidth]{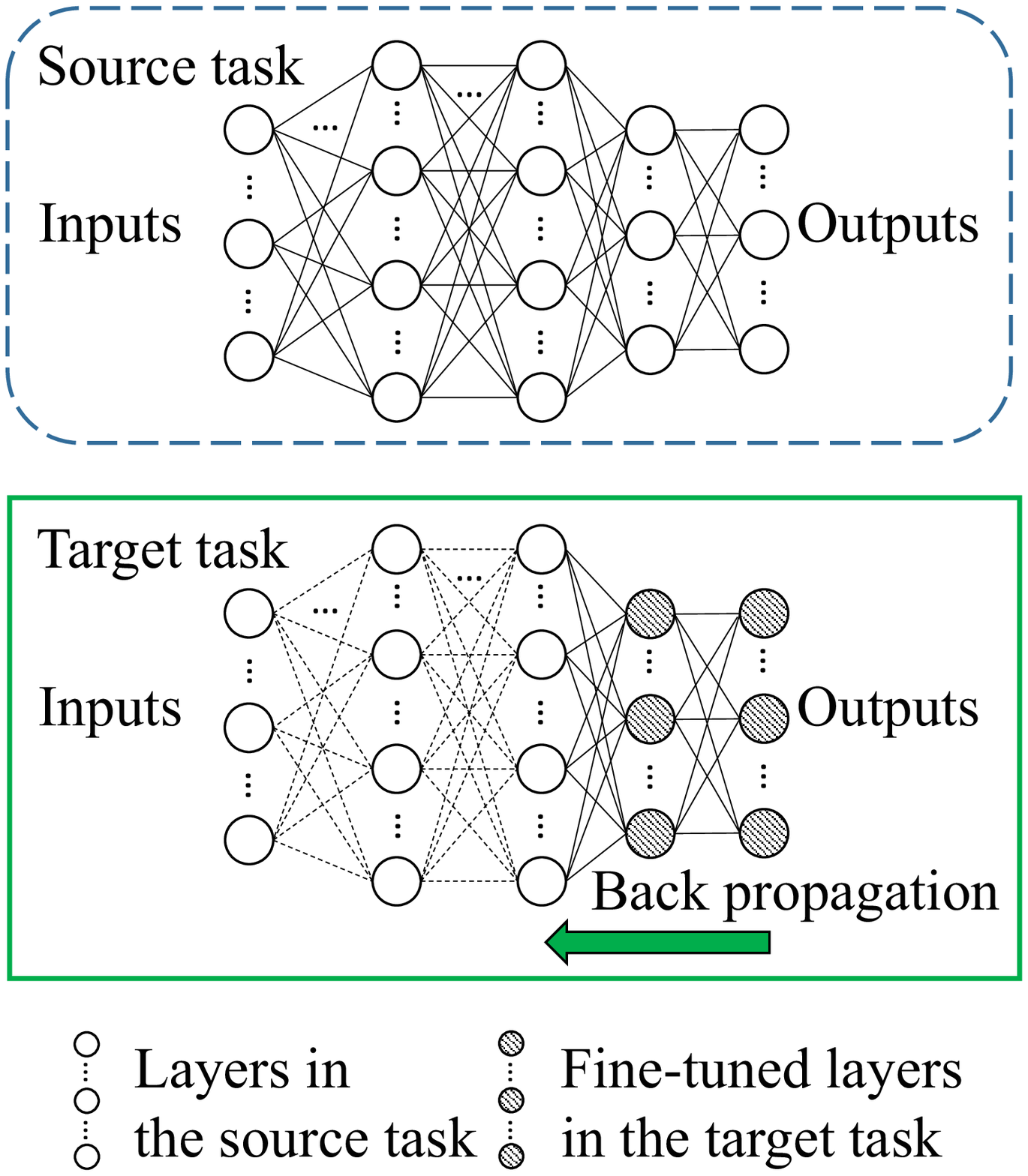}}
\subfigure[{Multiple types of services.}]{
\label{tf_plot} %% label for second subfigure
\includegraphics[width=0.35\textwidth]{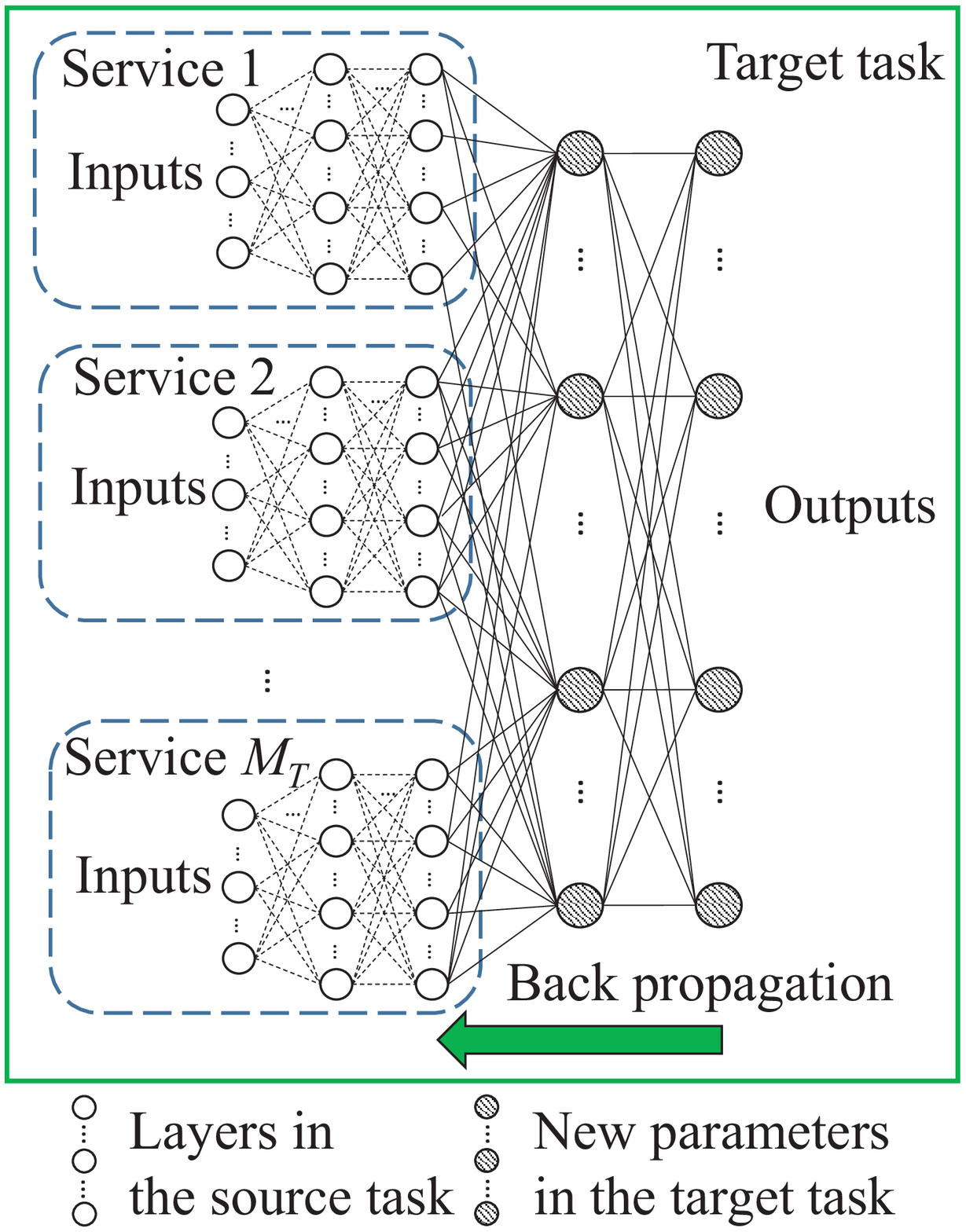}}
\caption{Deep transfer learning.}
 \label{fig:RD} %% label for entire figure
\vspace{-0.2cm}
\end{figure}

%\begin{figure}[htbp]
%	\vspace{-0.2cm}
%	\centering
%	\begin{minipage}[t]{0.4\textwidth}
%		\centering
%		\includegraphics[width=1\textwidth]{tf_one2one}
%	\end{minipage}
%	\vspace{-0.6cm}
%	\caption{Deep transfer learning with non-stationary wireless channels.}
%	\label{tf_one2one}
%	\vspace{-0.2cm}
%\end{figure}

\subsection{Transfer Learning with Non-Stationary Wireless Channels}
\label{change_channel}
In wireless communications, the distribution of small-scale channel gains may change over time. For example, the BS may switch ON/OFF some antennas. When the number of active antennas changes, the distribution of $g_k$ becomes different. For the cascaded NNs proposed in the previous section, the system fine-tunes the last a few layers of $\Phi_{\rm I}$ as illustrated in Fig. \ref{tf_one2one}. Since the power allocation policy depends on the distribution of wireless channels, the system fine-tunes all the layers of $\Phi_{\rm II}^{\xi}$.
%
%The source domain and the target domain with different channel distributions are denoted by $\mathcal{D}_{g}$ and $\mathcal{D}_{g'}$, respectively. The source task and the target task are denoted by $\mathcal{T}_{g}$ and $\mathcal{T}_{g'}$, respectively. Then, the deep transfer learning that exploits the knowledge obtained in the source domain and task in the target domain and task can be denoted by $\mathcal{D}_g, \mathcal{T}_g \rightarrow \mathcal{D}_{g'}, \mathcal{T}_{g'}$.
%
% needs to fine-tune the parameters of NNs in the second part since the power . For the NN in the first part, the input and output of the NN does not change with the distribution of wireless channels. As illustrated in , we can simply fine-tune the NN without changing its' structure.

\subsection{Transfer Learning with Different Types of Services}
In a wireless network, the service requests are highly dynamic. In other words, the number of different types of services in the wireless networks varies significantly over time. Thus, the system needs to update the NN according to the QoS requirements of different types of services. In the simulation, only one labeled training sample can be obtained in each epoch. Thus, the number of training samples used in transfer learning equals to the number of epochs it takes to converge.

%\begin{figure}[htbp]
%	\vspace{-0.2cm}
%	\centering
%	\begin{minipage}[t]{0.41\textwidth}
%		\centering
%		\includegraphics[width=1\textwidth]{tf_m2one}
%	\end{minipage}
%	\vspace{-0.6cm}
%	\caption{Deep transfer learning with multiple types of services.}
%	\label{tf_plot}
%	\vspace{-0.2cm}
%\end{figure}

\subsubsection{Transfer Learning from Delay-tolerant Services to Another Type of Services}
\label{singleservice}
For both delay-sensitive and URLLC services, the delay and reliability requirements depend on specific applications. Training NNs for all kinds of applications is not possible in practice. To overcome this difficulty, we first train a NN to approximate the optimal resource allocation policy of delay-tolerant services. Then, we fine-tune the NN for delay-sensitive and URLLC services. With the cascaded NNs in the previous section, the system needs to fine-tune the last a few layers of $\Phi_{\rm I}$ with the method in Fig. \ref{tf_one2one}. Since the power allocation policy depends on the QoS requirement of each service, all the layers of $\Phi^{\xi}_{\rm II}$ should be updated.

\subsubsection{Transfer Learning from a Single Type of Services to Multiple Types of Services}
\label{multiservice}
If there are $M_T$ types of services in the network, then there are $2^{M_T}$ possible combinations with different types of services. In 5G networks, $M_T$ will be large, and it is impossible to train a NN for each combination. If we have a well-trained NN for each type of services (i.e., source task), then by replacing the last a few layers of each NN, we can construct a new NN as that in Fig. \ref{tf_plot}. With the cascaded NNs, we only need to update $\Phi_{\rm I}$ for bandwidth allocation. The power allocation for each user is determined by $\Phi^{\xi}_{\rm II}$, which is the same as that in the source task. The algorithm is summarized in Table \ref{sub_tf}.

\begin{table}[hbt]
	\caption{Deep Transfer Learning for Multiple Types of Services}
	\vspace{-1.0cm}
	\begin{tabular}{p{17cm}}
		\\\hline
	\end{tabular}
	\label{sub_tf}
	\begin{algorithmic}[1]
\vspace{-0.5cm}
		\REQUIRE {Large-scale channel gains $\alpha_k^\xi$ and QoS constraints $c_k^\xi$.}
		\STATE Train a NN for delay-tolerant services.
        \STATE Initialize an empty set $\mathcal{S}_T$
		\FOR{the $m$-th type of services, $m \in  \{1, ..., M_T\}$}	
		\IF{the $m$-th type of services is requested by some users}
        \STATE $\mathcal{S}_T := \mathcal{S}_T \cup \{m\}$
		\STATE Collect a few training samples for the $m$-th type of services with the algorithm in Table \ref{sub2}.	
		\STATE Initialize parameters of a new NN with the parameters in the well-trained NN.
		\STATE Fine-tuning the last a few layers of the new NN with the method in Section \ref{singleservice}.
		\ENDIF
		\STATE Stack the NNs for all $m \in \mathcal{S}_T $ according to Fig. \ref{tf_plot}.
		\ENDFOR
		\STATE Collect a few training samples for multiple types of services with the algorithm in Table \ref{sub2}.
		\STATE Fine-tune the stacked NNs with the method in Section \ref{multiservice}.
		\RETURN {the parameters of the fine-tuned NNs.}
\vspace{-0.3cm}
	\end{algorithmic}
	\vspace{-0.3cm}
	\begin{tabular}{p{17cm}}
		\\\hline
	\end{tabular} \vspace{-0.2cm}
\end{table}

\section{Simulation and Numerical Results}
\label{simu}

In the considered scenario, the coverage of the BS is $200$ meters. Users are uniformly distributed around the BS. The path loss model is $35.3+37.6 \log_{10}(d) $, where $d$ is the distance (meters) between the BS and a user.
The shadowing is lognormal distributed with $8$ dB standard deviation. The small-scale channels are Rayleigh fading and the distribution of the small-scale channel gains follows $f_g(x) =
\frac{1}{(N_{\rm T}-1)!} x^{N_{\rm T}-1} e^{-x}$. The rest of the simulation parameters are summarized in Table \ref{tablesys}, unless specified otherwise.

\begin{table}[htbp]
	\centering
	\caption{Parameters in Simulation}
	\label{tablesys}
\vspace{-0.2cm}
	\begin{tabular}{|c|c|}%{|p{5.8cm}|p{2.2cm}|}
		\hline
		Maximal transmit power of the BS $P^{\max}$& $46$ dBm \\ \hline
		Duration of one TTI $T_{\rm s}$ & $0.125$ ms \cite{3GPP2017Agree}\\ \hline
		Bandwidth of each subcarrier $W$ & $120$ kHz \cite{3GPP2017Agree}\\ \hline
		Channel coherence time $T_{\rm c}$ & $5$ ms \cite{3GPP2017Agree}\\ \hline
		Single-sided noise spectral density $N_0$ & $-174$~dBm/Hz \\ \hline
		Number of bytes in a packet $B_k^{\rm u}$  & $[20,64]$ bytes \cite{3GPP2017Scenarios}\\ \hline
		Average data arrival rate of delay-tolerant users $\bar{a}_k$ &  $[50,100]$ KB/s \\ \hline
		Packet loss probability of URLLC $\epsilon^{\rm max,u}$ & $5 \times 10^{-8}$ \\ \hline
		Circuit power consumption per antenna $N^{\max}P^{\rm ca}$  & {$50$ mW} \cite{Debaillie2015PowerFormu} \\ \hline
		Fixed circuit power $P_0^{\rm c}$ & {$50$ mW} \cite{Debaillie2015PowerFormu} \\ \hline
		Power amplifier efficiency $\rho$  & $0.5$ \cite{Debaillie2015PowerFormu} \\ \hline
		Average packet arrival rate of delay-sensitive services $\nu^{\rm a}$  &  $[100, 1000]$ packets/s \\ \hline
		Average packet size of delay-sensitive services $1/\nu^{\rm s}$ & $[1, 20] ~\text{kbits} $ \\ \hline
		Delay bound of delay-sensitive user $D_k^{\rm q, s}$  & $50$ ms \\ \hline Maximal tolerable delay bound violation probability of delay-sensitive user $\epsilon_k^{\rm q, s}$ & $10^{-2}$ \\ \hline
	\end{tabular}
\end{table}
\vspace{-0.2cm}

\subsection{Validating the Properties of URLLC} % and Computing Approximation for URLLC services}
\begin{figure}[htbp]
	\vspace{-0.2cm}
	\centering
	\begin{minipage}[t]{0.7\textwidth}
		\centering
		\includegraphics[width=1\textwidth]{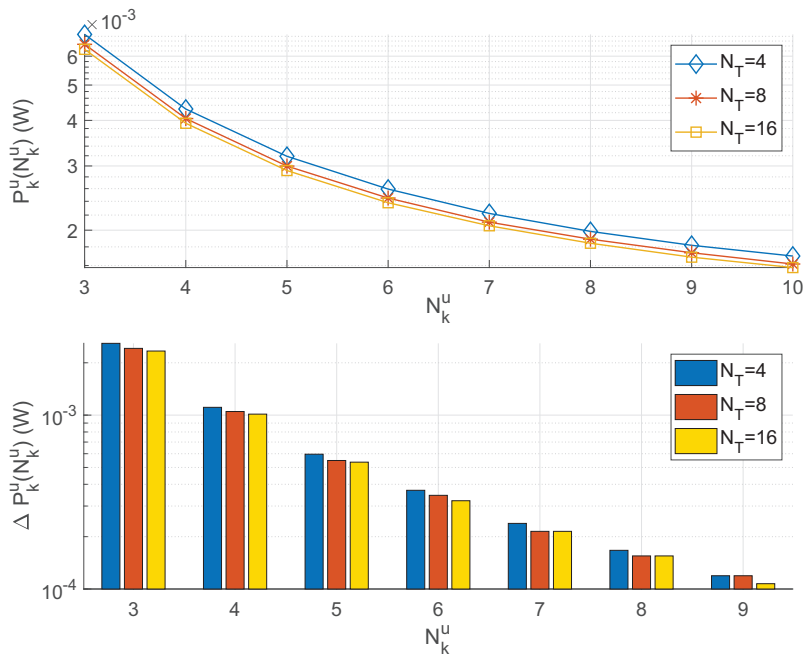}
	\end{minipage}
	\vspace{-0.6cm}
	\caption{Validating Conditions \ref{Condition1} and \ref{Condition2} for URLLC services.}
	\label{Fig:Con12}
	\vspace{-0.2cm}
\end{figure}

In this subsection, we first validate that Conditions \ref{Condition1} and \ref{Condition2} hold in non-asymptotic scenarios of URLLC. In Fig \ref{Fig:Con12}, we randomly select a user and illustrate the monotonicity of $P_k^{\rm u}(N_k^{\rm u})$ and $\Delta P_k^{\rm u}(N_k^{\rm u})$. The results show that even when the number of antennas is not large, such as $N_{\rm T} = 4, 8, 16$, Conditions \ref{Condition1} and \ref{Condition2} hold. The results indicate that the algorithms in Tables \ref{sub} and \ref{sub2} can converge to the optimal solutions.

\subsection{Performance Evaluation}
\label{deeplearningPerformance}
In this subsection, we evaluate the performance achieved by the deep learning method in Sections \ref{supervisedLearning} and \ref{sec:cascaded}. In the scenarios with multiple types of services, the ratio of the number of users requesting the three types of services is set to be $1:1:1$. The algorithm in Table \ref{sub2} is used to find the optimal solutions of problem \eqref{obj_tot} with $10000$ inputs, where $N^{\max}= 256$, $N_T=64$ and $K^\xi=20$. The first $9000$ samples are used to train the NNs and the last $1000$ samples are used to test the performance of them. In each epoch, $M_t = 128$ training samples are randomly selected from $9000$ training samples, and the learning rate is set to be $0.001$. The DL algorithm is implemented in Python with TensorFlow 1.11.

%
%
%The number of training samples in each epoch is $M_t = 128$ and the memory can save up to $1024$ training samples \cite{angela2018DQN}. We set the learning rate of the DNN as $0.001$.

Each neural network consists of one input layer, one output layer, and ${L}_{\text{hidden}}^\xi$ hidden layers, where each hidden layer has $N_{\text{neurons}}^\xi$ neurons. The input and output layers of the FNN are defined after \eqref{eq:opt}. The input and output layers of the cascaded NNs are defined in Fig. \ref{cas}. The hyper-parameters (i.e., ${L}_{\text{hidden}}^\xi$ and $N_{\text{neurons}}^\xi$) for different types of services can be found in Table \ref{HyperParameter}. We selected the hyper-parameters by trial and error, where ${L}_{\text{hidden}}^\xi$ ranges from $1$ to $10$ and $N_{\text{neurons}}^\xi$ ranges from $200$ to $1000$. The hyper-parameters in Table \ref{HyperParameter} can achieve the best performance according to our experience.

\begin{table}[htbp]
	\centering
	\caption{Hyper-parameters of NNs}
	\label{HyperParameter}
\vspace{-0.2cm}
	\begin{tabular}{|c|c|c|c|c|c|c|}
		\hline
		\multirow{3}*{Service type}  & \multicolumn{2}{|c|}{\multirow{2}*{FNN}} & \multicolumn{4}{|c|}{Cascaded NNs}\\
		\cline{4-7}
		\multicolumn{1}{|c|}{~} & \multicolumn{2}{|c|}{~} & \multicolumn{2}{|c|}{The $1$st part $\Phi_{\rm I}$} & \multicolumn{2}{|c|}{The $2$nd part $\Phi^{\xi}_{\rm II}$} \\
        \cline{2-7}
        \multicolumn{1}{|c|}{~} & ${L}_{\text{hidden}}^\xi$ & $N_{\text{neurons}}^\xi$ & ${L}_{\text{hidden}}^\xi$ & $N_{\text{neurons}}^\xi$ & ${L}_{\text{hidden}}^\xi$ & $N_{\text{neurons}}^\xi$ \\
		\hline
        Delay-tolerant & 4 & 800 & 4 & 800 & 4 & 20\\\hline
        Delay-sensitive & 5 & 600 & 5 & 600& 4 & 20 \\\hline
        URLLC & 4 & 600 & 4 & 600& 4 & 20 \\\hline
	\end{tabular}
\end{table}
\vspace{-0.0cm}

\begin{figure}[htbp]
	\vspace{-0.4cm}
	\centering
	\begin{minipage}[t]{0.7\textwidth}
		\centering
		\includegraphics[width=0.9\textwidth]{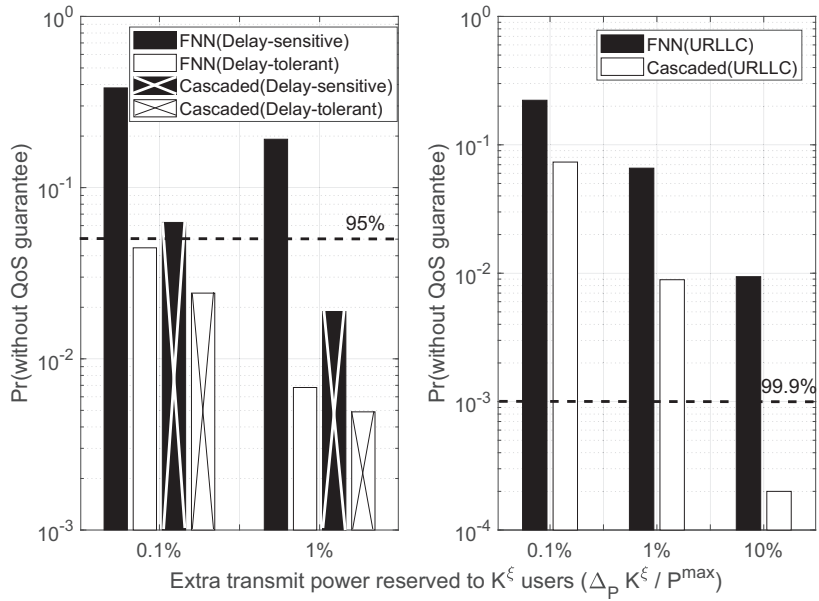}
	\end{minipage}
	\vspace{-0.2cm}
	\caption{Probability without QoS guarantee v.s. extra transmit power reserved to the users.}
	\label{qos}
	\vspace{-0.4cm}
\end{figure}

In Fig. \ref{qos}, we show the QoS achieved by the FNN and the cascaded NNs. Specifically, the relation between the probability without QoS guarantee (i.e., the probability that the transmit power allocated to a user is smaller than the minimum transmit power that is required to satisfy the QoS constraint of the user, $\Pr\{ \Phi_{\rm II}^{\xi}(\boldsymbol{X}_k^{\xi},\varLambda_{\rm II}^{\xi})+\Delta_P < P_k^{\xi}(\tilde{N}_k^{\xi} ) \}$). and the extra transmit power reserved to all the $K^\xi$ users, $\Delta_P K^\xi$, is provided. For each type of service, we set $|\mathcal{K}^\xi|=20$. The results are evaluated with $1000$ testing samples.

%We show whether the learned $\tilde{\boldsymbol{P}}$ and $ \tilde{\boldsymbol{N}}$ in All-in-One and cascaded structures' outputs $\tilde{\boldsymbol{Y}}$ can guarantee the QoS as discussed in Section \ref{DLproblem} or not. By counting how many users' QoS can be met by the learned $\tilde{\boldsymbol{P}}$ and $ \tilde{\boldsymbol{N}}$ in the test samples, and calculating the ratio of the unsatisfied number of users to the total number of users, which is $K^\xi=20$, we can quantify the probability without QoS guarantee. To improve the probability, we reserve a minor amount of the total transmit power $P^{\max}$ to compensate the approximation error from the outputs of neural networks. Without loss of generosity, this compensation is evenly assigned to all the users.
%The result in Fig. \ref{qos} are averaged over $1000$ samples.
%
From Fig. \ref{qos}, we can observe that the cascaded NNs can achieve better QoS compared with the FNN. For example, by reserving $10$\% of $P^{\max}$ extra transmit power to the $20$ URLLC users, the cascaded NNs can satisfy the QoS requirement with a probability of $99.98\%$. However, the FNN can only satisfy the QoS requirement with a probability of $99.2$\%. For other types of services, the cascaded NNs also outperform the FNN in terms of achieving better QoS. This validates that the cascaded NNs can improve the QoS for all types of services.

\begin{figure}[htbp]
	\vspace{-0.4cm}
	\centering
	\begin{minipage}[t]{0.6\textwidth}
		\centering
		\includegraphics[width=1\textwidth]{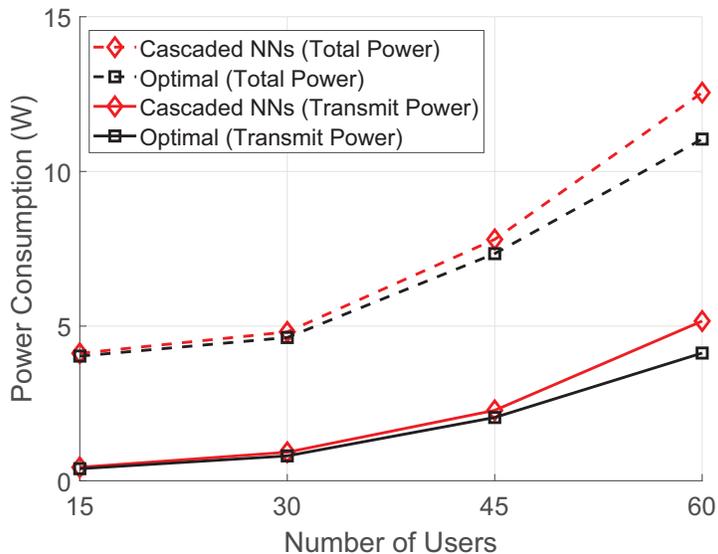}
	\end{minipage}
	\vspace{-0.2cm}
	\caption{Power consumption v.s. number of users when $N_T = 64$ and $N^{\max}= 256$. }
	\label{lessuser}
	\vspace{-0.4cm}
\end{figure}

The total power consumption and transmit power achieved with different schemes are illustrated in Fig.~\ref{lessuser}. We compare the performance of the cascaded NNs with the optimal solutions obtained with the algorithm in Tables~\ref{sub} and \ref{sub2} (with legend `Optimal').
%The first one is an existing approach that use game theory to design resource allocation (with legend `Game')\footnote{Since there is no existing method that can guarantee the QoS of different kinds of services, we include constraints \eqref{dtCon}, \eqref{dsCon}, and \eqref{err} in the game theory problem in \cite{Sacchi2011Game} to get this benchmark.} \cite{Sacchi2011Game}.
%The second one is the minimum power consumption with the optimal solutions obtained with the algorithm in Tables~\ref{sub} and \ref{sub2} (with legend `Optimal').
For the deep learning method, we train the cascaded NNs when $K^{\rm t} = K^{\rm s} = K^{\rm u} =20$, which is close to the maximal number of users that can be served with the given radio resources\footnote{The maximal number of users that can be served by a BS depends on the distribution of the users. We set the user-BS distance equals to the radium of the cell to calculate the maximal number of users.}. In practical systems, the number of users is dynamic. When the number of users is less than $60$, we do not change the dimension of the input, but set $c_k^{\xi} = 0$. It means that the required data rates, effective bandwidth or packet sizes of some users are zero. In this case, no resource will be assigned to them.
%The results in Fig.~\ref{lessuser} show that the cascaded NNs outperform `Game', especially when the number of user is large. In addition,
The performance with the cascaded NNs is close to the optimal solutions. This implies the cascaded NNs are a good approximation of the optimal policy.

\subsection{Performance with Transfer Learning}
Since NNs are used to approximate the optimal resource allocation policy, the accuracy is defined as follows,
\begin{equation}
\eta = 1 - \text{Error} =
1- \frac{ {P}_{\rm tot}(\tilde{\boldsymbol{N}}, \tilde{\boldsymbol{P}}) -  P_{\rm tot}(\boldsymbol{N}^*, \boldsymbol{P}^*)}
{P_{\rm tot}(\boldsymbol{N}^*,\boldsymbol{P}^*)},
\end{equation}
which reflects the gap between the outputs of NNs and the optimal solutions.

To show that convergence time of different methods, we provide the relation between the numbers of training epochs and the accuracy. The transfer learning methods that fine-tune the well-trained NNs in the source domain and task are compared with the benchmark that trains new NNs with randomly initialized parameters (with legend `Random initialization' and initializing each parameter with a zero mean and unit variance Gaussian variable). In this subsection, we only consider the cascaded NNs since this structure can guarantee the QoS constraints with a high probability.

%The training losses throughout the entire paper are plotted every $10$ learning epochs;
%he accuracies of all kinds of services throughout the entire paper are averaged over the last $10$ epochs.

\begin{figure*}[htbp]
	\vspace{-0.4cm}
	\centering
	\begin{minipage}[t]{1\textwidth}
		\centering
		\includegraphics[width=1\textwidth]{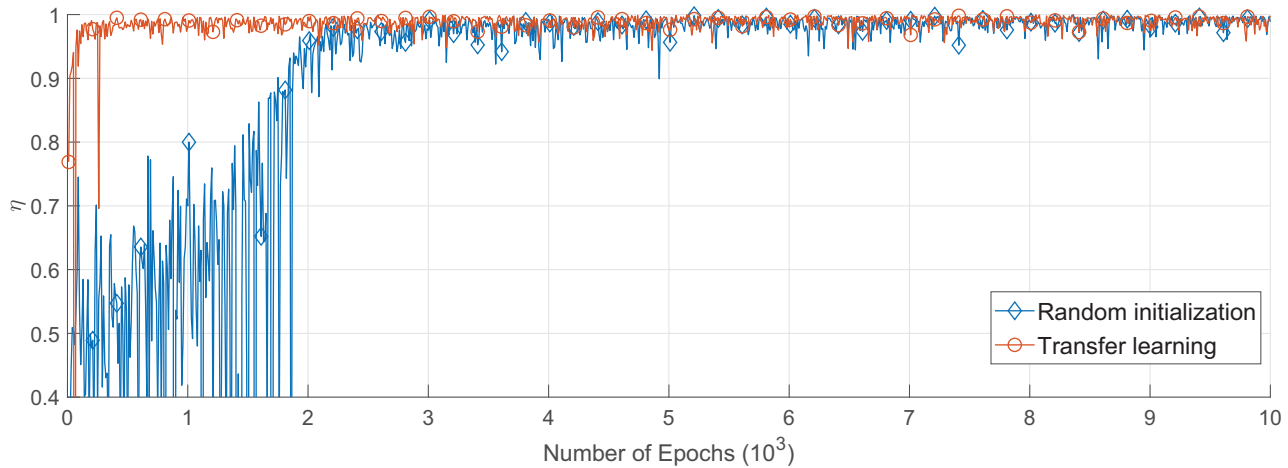}
	\end{minipage}
	\vspace{-0.9cm}
	\caption{Accuracy v.s. the number of training epochs when $N_T$ varies, where $N^{\max}= 256$, $K^{\rm u}=20$.}
	\label{Nt16-Nt64}
	\vspace{-0.4cm}
\end{figure*}

\subsubsection{Transfer learning with non-stationary wireless channels} The training samples in the source domain and task are obtained when $N_{\rm T} = 16$. The training samples in the target domain and task are obtained when  $N_{\rm T} = 64$. With different numbers of antennas, the distribution of small-scale channel gains varies. With transfer learning, the first $3$ layers of $\Phi_{\rm I}$ are fixed. The last layer of $\Phi_{\rm I}$ and $\Phi^{\xi}_{\rm II}$ are fine-tuned. The results in Fig. \ref{Nt16-Nt64} show that with transfer learning, only $400$ epochs ($400$ training samples in the new scenario) are needed to achieve around \blue{$0.98$} accuracy, while $2000$ epochs ($2000$ training samples in the new scenario) are needed to achieve the same accuracy with random initialization.

\begin{figure*}[htbp]
	\vspace{-0.4cm}
	\centering
	\begin{minipage}[t]{1\textwidth}
		\centering
		\includegraphics[width=1\textwidth]{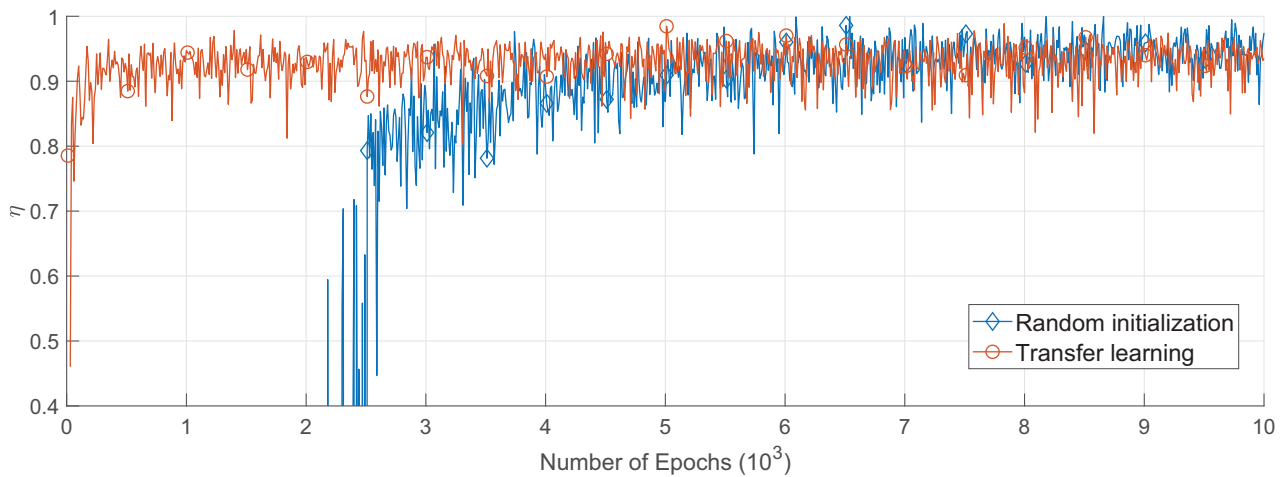}
	\end{minipage}
	\vspace{-0.9cm}
	\caption{Accuracy v.s. the number of training epochs, where the target task is resource allocation for delay-sensitive services, $N^{\max}= 256$, $N_T=64$ and $K^\xi=20$.}
	\label{DT-DS}
	\vspace{-0.4cm}
\end{figure*}

\begin{figure*}[htbp]
	\vspace{-0.2cm}
	\centering
	\begin{minipage}[t]{1\textwidth}
		\centering
		\includegraphics[width=1\textwidth]{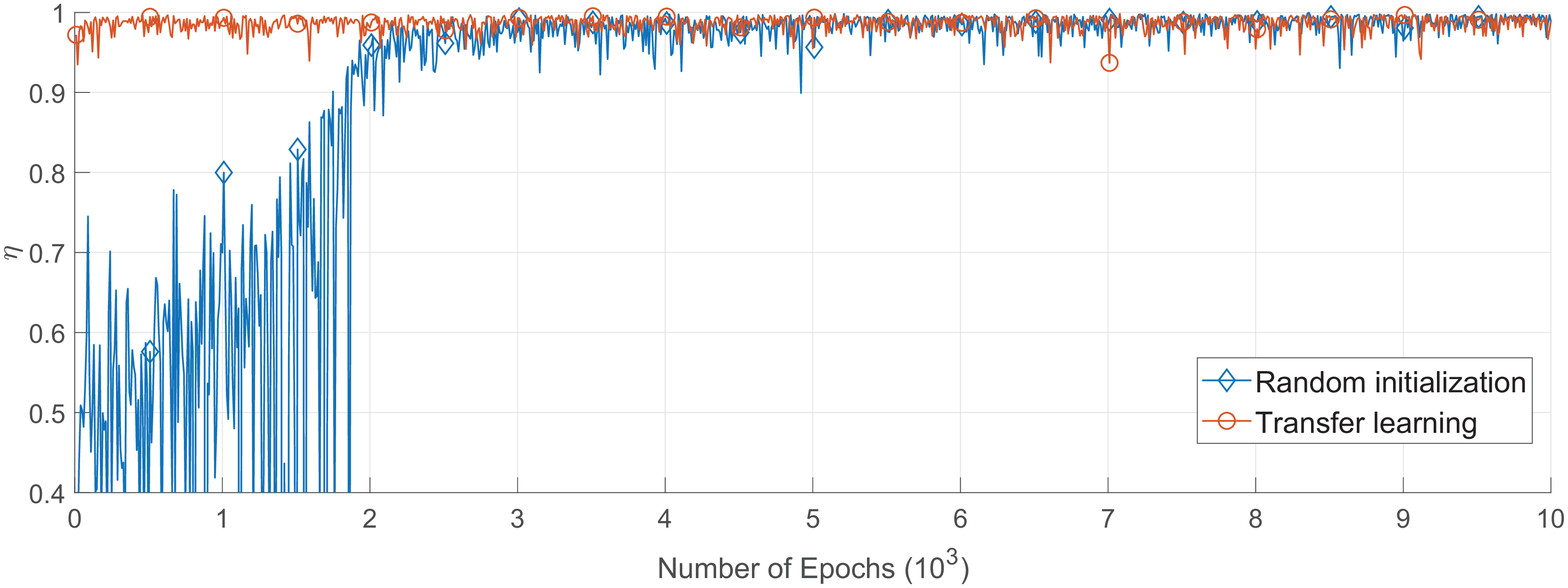}
	\end{minipage}
	\vspace{-0.9cm}
	\caption{Accuracy v.s. the number of training epochs, where the target task is resource allocation for URLLC, $N^{\max}= 256$, $N_T=64$ and $K^\xi=20$.}
	\label{DT-U}
	\vspace{-0.4cm}
\end{figure*}

\subsubsection{Transfer learning from delay-tolerant services to another type of services}
\label{singletosingle} We first train cascaded NNs for delay-tolerant services with $9000$ labeled training samples. Then, we fine-tune the well-trained cascaded NNs with new labeled training samples of another type of services. Specifically, the first $3$ layers of $\Phi_{\rm I}$ are fixed and the NNs in the second part, $\Phi_{\rm II}^{\rm t}$, are replaced with NNs for delay-sensitive services, $\Phi_{\rm II}^{\rm s}$.

The results in Fig. \ref{DT-DS} show that the transfer learning method can achieve $0.9$ accuracy with around $150$ epochs ($150$ training samples in the new scenario), while it takes $2500$ epochs for the random initialization method to achieve the same accuracy ($2500$ training samples in the new scenario). A similar conclusion can be observed from the results in Fig. \ref{DT-U}. By comparing the results in Figs. \ref{DT-DS} and \ref{DT-U}, we can see that the accuracy of transfer learning for URLLC is higher than that for delay-sensitive services. As shown in Table \ref{HyperParameter}, to achieve good performance for delay-sensitive services, we need $5$ hidden-layers in $\Phi_{\rm I}$. However, for delay-tolerant and URLLC services, only $4$ hidden-layers are needed in $\Phi_{\rm I}$. Since we use the same hyper-parameter in the source task and target tasks, deep transfer learning achieves higher accuracy for URLLC compared with delay-sensitive services.

\begin{figure*}[htbp]
	\vspace{-0.2cm}
	\centering
	\begin{minipage}[t]{\textwidth}
		\centering
		\includegraphics[width=\textwidth]{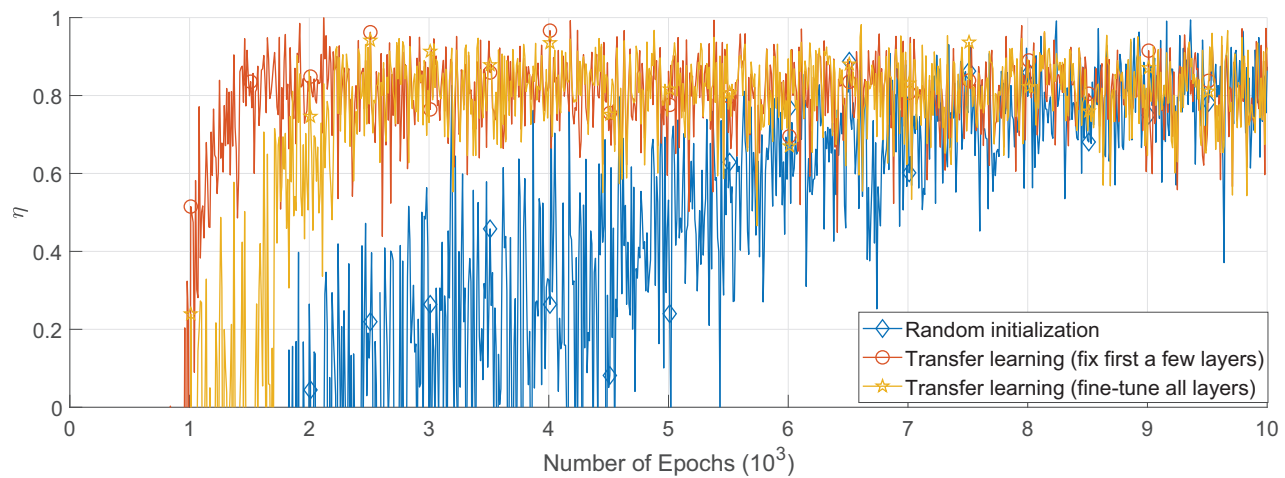}
	\end{minipage}
	\vspace{-1.0cm}
	\caption{Transfer knowledge from networks with a single type of services to networks with multiple types of services, $N^{\max}= 256$, $N_T=64$ and $K^\xi=20$.}
	\label{multi-service}
	\vspace{-0.2cm}
\end{figure*}

\subsubsection{Transfer knowledge from a single type of services to multiple types of services}
To apply transfer learning in bandwidth allocation, the structure in Fig. \ref{tf_plot} is adopted. Specifically, the output layers of the three NNs for the three types of services are replaced with an output layer with $(K^{\rm t}+K^{\rm s}+K^{\rm u})$ neurons. With deep transfer learning, we can either fix the first a few-layers or fine-tune all the layers, i.e., the curves with legends `Transfer learning (fix first a few layers)' and `Transfer learning (fine-tune all layers)', respectively. For the neural network with legend `Transfer learning (fix first a few layers)', we fixed the first a few layers and fine-tuned the last $2$ layers. The performance of them is compared with a benchmark that trains a NN with randomly initialized parameters (with legend `Random initialization'), where the NN includes $4$ hidden layers, and each of them has $800$ neurons. The results in Fig. \ref{multi-service} show that `Transfer learning (fix first a few layers)' outperforms `Transfer learning (fine-tune all layers)' in the first $2500$ epochs (with less than $2500$ training samples), and they achieve the same performance after the $2500$ epoch (with more than $2500$ training samples). This indicates that there is no need to fine-tune all the layers of the NN. Compared with the benchmark, transfer learning can achieve higher accuracy in the first $8000$ training epochs. By the end of the training phase, the performance of them is almost the same.

\section{Conclusion}
In this work, we studied how to use deep learning in resource allocation with diverse QoS requirements in 5G networks. Specifically, we proposed an optimization algorithm that can converge to the optimal solution of an optimization problem that minimizes the total power consumption for delay-tolerant, delay-sensitive, and URLLC services. The obtained optimal solutions were used as labeled training samples to train NNs that approximate the optimal policy. To guarantee the diverse QoS requirements in non-stationary wireless networks, we designed cascaded NNs and fine-tuned their parameters with deep transfer learning. Our simulation results validated that the proposed deep transfer learning framework converges quickly when the wireless channels or the service requests are non-stationary.

\appendices

\section{Proof of Optimality of the Algorithm in Table \ref{sub}}
\label{App:Prop1}
\renewcommand{\theequation}{A.\arabic{equation}}
\setcounter{equation}{0}
\begin{proof}
We denote the objective function in \eqref{obj} as $f(\boldsymbol{N}) = \sum_{ k\in \cal K^{\xi} } P_k^{\xi}(N_k^{\xi})$ in this Appendix, where $\boldsymbol{N} = [N_1^{\xi}, ..., N_K^{\xi}]^{\rm T}$. The outcome of the algorithm in Table \ref{sub} is denoted by $\boldsymbol{N}^* = [N_1^{\xi*}, ..., N_K^{\xi*}]^{\rm T}$. To prove the optimality of the proposed algorithm, we only need to prove that for any bandwidth allocation scheme $\boldsymbol{N}' = [N_1^{\xi'}, ..., N_K^{\xi'}]^{\rm T}$, $f(\boldsymbol{N}^*) \leq f(\boldsymbol{N}')$ holds.

The difference between 	$\boldsymbol{N}^*$ and $\boldsymbol{N}'$ is denoted by	
\begin{align} \label{deltaN}
	\Delta \boldsymbol{N}
	= \boldsymbol{N}' - \boldsymbol{N}^* = [ \Delta N_1, \Delta N_2, ..., \Delta N_K ]^{\rm T}.
\end{align}

We further denote that
\begin{align}
\boldsymbol{N}^+ = [ \max(0, \Delta N_1),..., \max(0, \Delta N_K)],\nonumber\\
\boldsymbol{N}^- = [ -\min(0, \Delta N_1),..., -\min(0, \Delta N_K)].\nonumber
\end{align}
Then, we can obtain a bandwidth allocation policy $\boldsymbol{N}^0 = \boldsymbol{N}^* - \boldsymbol{N}^- = \boldsymbol{N}' - \boldsymbol{N}^+$.

The required transmit power with policy $\boldsymbol{N}^0 = [N^0_1,...,N^0_K]$ is $f(\boldsymbol{N}^0)$. Based on $\boldsymbol{N}^0$, if we allocate $(N^{\max} - \sum_{k=1}^{K}N_k^0)$ extra subcarriers to the users according to $\boldsymbol{N}^+$, the amount of power saving is $f(\boldsymbol{N}^0) - f(\boldsymbol{N}')$. If we allocate $(N^{\max} - \sum_{k=1}^{K}N_k^0)$ subcarriers according to $\boldsymbol{N}^-$, then the amount of power saving is $f(\boldsymbol{N}^0) - f(\boldsymbol{N}^*)$.

The amount of power saving with the above two approaches can be expressed as the sum of $(N^{\max} - \sum_{k=1}^{K}N_k^0)$ terms, i.e.,
\begin{align}
&f(\boldsymbol{N}^0) - f(\boldsymbol{N}') = \sum_{ k^+ \in \cal K^{+} }  \sum_{n = N_{k^+}^0}^{N_{k^+}^{'} - 1}  \Delta P_{k^+}(n),\label{eq:saving1}\\
&f(\boldsymbol{N}^0) - f(\boldsymbol{N}^*) = \sum_{ k^- \in \cal K^{-} }  \sum_{n = N_{k^-}^0}^{N_{k^-}^* - 1}  \Delta P_{k^-}(n),\label{eq:saving2}
\end{align}
where $\mathcal{K}^+ = \{ k | \Delta N_k >0 \}$ and $\mathcal{K}^- = \{ k | \Delta N_k <0 \}$.

According to Condition \ref{Condition1} and \ref{Condition2}, we have
\begin{align}
\Delta P_{k^+}(N_{k^+}^0) &\geq \Delta P_{k^+}(N_{k^+}^0 + 1) \geq ... \geq \Delta P_{k^+}(N_{k^+}^{'} - 1),\forall k^+ \in \mathcal{K}^+.\label{+deltaP}\\
\Delta P_{k^-}(N_{k^-}^0) &\geq \Delta P_{k^-}(N_{k^-}^0 + 1) \geq ... \geq \Delta P_{k^-}(N_{k^-}^* - 1), \forall k^- \in \mathcal{K}^-.
	\label{-deltaP}	
\end{align}
With the proposed algorithm, a subcarrier will be assigned to the user with the highest power saving. Thus, we have
	\begin{equation}
	\Delta P_{k^-}(N_{k^-}^* - 1)  \geq \Delta P_{k^+}(N_{k^+}^0), \forall k^+ \in \mathcal{K}^+, \forall k^- \in \mathcal{K}^-.
	\end{equation}
Since $\Delta P_{k^-}(N_{k^-}^* - 1)$ is the last term in \eqref{-deltaP} and $P_{k^+}(N_{k^+}^0)$ is the first term in \eqref{+deltaP}, we can obtain that each term in the right-hand side of \eqref{eq:saving1} is smaller than any term in the right-hand side of \eqref{eq:saving2}. Therefore, we have $f(\boldsymbol{N}^0) - f(\boldsymbol{N}') \leq f(\boldsymbol{N}^0) - f(\boldsymbol{N}^*)$, and hence $f(\boldsymbol{N}^*) \leq f(\boldsymbol{N}')$. This completes the proof.
\end{proof}

\section{Proof of Property \ref{P2}}
\label{App:Prop2}
\renewcommand{\theequation}{B.\arabic{equation}}
\setcounter{equation}{0}
\begin{proof}
	For notational simplicity, we replace $\mathord{\buildrel{\lower3pt\hbox{$\scriptscriptstyle\smile$}} \over f} _k^\xi(P_k^\xi,\mathord{\buildrel{\lower3pt\hbox{$\scriptscriptstyle\smile$}} 	\over N} _k^\xi )$ with $f_k^\xi( P_k^\xi, N_k^\xi )$ in this appendix.
We first derive the first-order derivatives of the right-hand and the left-hand sides of $f_k^\xi( P_k^\xi, N_k^\xi ) = c^{\xi}_k$, i.e.,
\begin{equation}
\label{f_first}
\frac{\partial f_k^\xi( P_k^\xi, N_k^\xi )}
{\partial N_k^\xi}  +
\frac{\partial f_k^\xi( P_k^\xi, N_k^\xi )}
{\partial P_k^\xi}
\frac{\partial  P_k^\xi}{\partial N_k^\xi} = 0.
\end{equation}
Since $f_k^\xi( P_k^\xi, N_k^\xi )$ increases with both $N_k^\xi$ and $P_k^\xi$, we have $\frac{\partial f_k^\xi( P_k^\xi, N_k^\xi )}	{\partial N_k^\xi} > 0$ and $\frac{\partial f_k^\xi( P_k^\xi, N_k^\xi )}
{\partial P_k^\xi} >0 $. According to \eqref{f_first}, we can see that $\frac{\partial  P_k^\xi}{\partial N_k^\xi} < 0$, i.e., $P_k^\xi$ decreases with $N_k^\xi$. Therefore, Condition \ref{Condition1} holds.

%We know that the joint concave function $f_k^\xi( P_k^\xi, N_k^\xi )$ is an increasing function of both $N_k^\xi$ and $P_k^\xi$. That are $\frac{\partial f_k^\xi( P_k^\xi, N_k^\xi )}	{\partial N_k^\xi} > 0$ and $\frac{\partial f_k^\xi( P_k^\xi, N_k^\xi )}
%{\partial P_k^\xi} >0 $. To hold the equality of \eqref{f_first}, we can determine that $P_k^\xi$ decreases with $N_k^\xi$, i.e., $\frac{\partial  P_k^\xi}{\partial N_k^\xi} < 0$. This conclusion meets the requirement of Condition \ref{Condition1}.

From \eqref{f_first}, we further derive the second-order derivative, i.e.,
\begin{align}
\label{f_second}
&\underbrace{\frac{\partial^2 f_k^\xi( P_k^\xi, N_k^\xi )}
	{\partial (P_k^\xi)^2}}_{a}
\underbrace{\left(\frac{\partial P_k^\xi}
	{\partial N_k^\xi} \right)^2}_{x^2}
+
2 \underbrace{\frac{\partial^2 f_k^\xi( P_k^\xi, N_k^\xi )}
	{\partial N_k^\xi \partial P_k^\xi} }_{b}
\underbrace{\frac{\partial P_k^\xi}
	{\partial N_k^\xi} }_{x}
+ \underbrace{ \frac{\partial^2 f_k^\xi( P_k^\xi, N_k^\xi )}
	{\partial (N_k^\xi)^2} }_{c}
+
\underbrace{\frac{\partial f_k^\xi( P_k^\xi, N_k^\xi )}
	{\partial P_k^\xi} \frac{\partial^2 P_k^\xi}
	{\partial (N_k^\xi)^2}}_{d}
= 0.
\end{align}
For notational simplicity, we can simplify \eqref{f_second} as $ax^2+2bx+c+d = 0$, which can be re-expressed as follows,
\begin{align}
\label{f_second_simp}
a \left[ \left(x+\frac{b}{a}\right)^2 + \frac{ac-b^2}{a^2} \right] + d = 0.
\end{align}
Since $f_k^\xi( P_k^\xi, N_k^\xi )$ is jointly concave in $P_k^\xi$ and $N_k^\xi$, we have $a = \frac{\partial^2 f_k^\xi( P_k^\xi, N_k^\xi )} {\partial (P_k^\xi)^2} \leq 0$ and $ \frac{\partial^2 f_k^\xi( P_k^\xi, N_k^\xi )} {\partial (P_k^\xi)^2}  \frac{\partial^2 f_k^\xi( P_k^\xi, N_k^\xi )} {\partial (N_k^\xi)^2} $ $- \left(\frac{\partial^2 f_k^\xi( P_k^\xi, N_k^\xi )}
{\partial N_k^\xi \partial P_k^\xi}\right)^2 \geq 0 $, i.e., $ac-b^2\geq 0$. Thus, from \eqref{f_second_simp}, we can see that $d =  \frac{\partial f_k^\xi( P_k^\xi, N_k^\xi )}
	{\partial P_k^\xi} \frac{\partial^2 P_k^\xi}
	{\partial (N_k^\xi)^2} \geq 0$. Further considering that $\frac{\partial f_k^\xi( P_k^\xi, N_k^\xi )}	{\partial P_k^\xi} >0 $, we can conclude that $\frac{\partial^2 P_k^\xi}
{\partial (N_k^\xi)^2} \geq 0$, i.e., $P_k^{\xi}$ is convex in $N_k^{\xi}$. Therefore, Condition \ref{Condition2} holds. The proof follows.
\end{proof}

\section{Proof of Optimality of the Algorithm in Table \ref{sub2} }
\label{App:Prop3}
\renewcommand{\theequation}{C.\arabic{equation}}
\setcounter{equation}{0}
\begin{proof}	
We denote the objective function in \eqref{obj_tot} as $f(\boldsymbol{N}) = \frac{1}{\rho} \sum_{ k\in \cal K^{\xi} }P_k^\xi(N_k^\xi) + P^{\rm ca}N_{\rm T} \sum_{ k\in \cal K^{\xi} }N_k^\xi + P_0^{\rm c}$ in this Appendix, where $\boldsymbol{N} = [N_1^{\xi}, ..., N_K^{\xi}]^{\rm T}$. The bandwidth allocation obtained in Line 10 and Line 12 (or 21) in Table \ref{sub2} are denoted by $\boldsymbol{\check{N}} = [\check{N}_1^{\xi}, ..., \check{N}_K^{\xi}]^{\rm T}$ and $\dot{\boldsymbol{N}} = [\dot{N}_1^{\xi}, ..., \dot{N}_K^{\xi}]^{\rm T}$, respectively.

Since the algorithm in Lines 2-10 in Table \ref{sub2} is similar to the algorithm in Table \ref{sub}, with the method in Appendix \ref{App:Prop1}, we can prove that $\boldsymbol{\check{N}} = [\check{N}_1^{\xi}, ..., \check{N}_K^{\xi}]^{\rm T}$ minimizes $f(\boldsymbol{N})$ when Conditions \ref{Condition1} and \ref{Condition2} hold.

If $\sum_{ k\in \cal K^{\xi} } \check{P}_k^{\xi}(\check{N}_k^{\xi}) \leq  P^{\max}$, then the resource allocation satisfies the transmit power constraint, and $\check{N}_k^{\xi}$ and $\check{P}_k^{\xi}(\check{N}_k^{\xi})$, $k=1,...,K,$ are the optimal solution of problem \eqref{obj_tot}.

If $\sum_{ k\in \cal K^{\xi} } \check{P}_k^{\xi}(\check{N}_k^{\xi}) >  P^{\max}$, the resource allocation that minimizes the total power consumption does not satisfies the maximal transmit power constraint. From the algorithm in Table \ref{sub}, we know whether problem \eqref{obj_tot} is feasible or not. In the cases that the problem is feasible, we have $\sum_{ k\in \cal K^{\xi} } \check{N}_k^{\xi} <  N^{\max}$ when $\sum_{ k\in \cal K^{\xi} } \check{P}_k^{\xi}(\check{N}_k^{\xi}) >  P^{\max}$. From the condition in Line 4 in Table \ref{sub2}, we know that $\Delta P_{\mathrm{tot}, k}^{\xi}(\check{N}_{k}^{\xi})>0, \forall k = 1,...,K$. Thus, the total power consumption increases with ${N}_{k}^{\xi}$ when ${N}_{k}^{\xi} \geq \check{N}_{k}^{\xi}$. Minimizing the total power consumption is equivalent to minimizing the number of subcarriers that can guarantee the maximal transmit power constraint. In the rest part of this appendix, we prove that the algorithm in Table \ref{sub2} can find the minimal number of subcarriers.

The algorithm from Lines 15-20 in Table \ref{sub2} is the same as that in Table \ref{sub}. Thus, the bandwidth allocation obtained in each iteration minimizes the sum of the required transmit power. According to the condition in Line 15 in Table \ref{sub2}, if the total number of occupied subcarriers is less than $\sum_{ k\in \cal K^{\xi} } \dot{N}_k^{\xi}$, then the maximal transmit power constraint cannot be satisfied. Therefore, $\sum_{ k\in \cal K^{\xi} } \dot{N}_k^{\xi}$ is the minimum number of subcarriers that is required to satisfy the maximal transmit power constraint. This completes the proof.

\end{proof}

\bibliographystyle{IEEEtran}
\bibliographystyle{unsrt} % in order of presenting sequence
\bibliography{bibfile}

\end{document}